\documentclass[10pt,twocolumn,letterpaper]{article}

\usepackage{cvpr}      
\definecolor{cvprblue}{rgb}{0.21,0.49,0.74}
\usepackage[pagebackref,breaklinks,colorlinks,allcolors=cvprblue]{hyperref}

\def\paperID{*****} 
\def\confName{CVPR}
\def\confYear{2026}

\title{MeshTailor: Cutting Seams via Generative Mesh Traversal}

\author{
    Xueqi Ma\textsuperscript{1} \quad
    Xingguang Yan\textsuperscript{2} \quad
    Congyue Zhang\textsuperscript{1} \quad
    Hui Huang\textsuperscript{1*}\\[0.2em]
    \textsuperscript{1}Shenzhen University \quad
    \textsuperscript{2}Simon Fraser University
}

\usepackage{color}
\usepackage[table]{xcolor}

\usepackage{marvosym}

\let\digamma\relax

\usepackage{amsthm, amsfonts, amssymb}
\usepackage{makecell}
\usepackage{textcomp}
\usepackage{epstopdf}
\usepackage{multirow}
\usepackage{wrapfig}
\usepackage[ruled]{algorithm2e} 
\usepackage{bm}
\usepackage{enumitem}

\SetAlFnt{\small}
\SetAlCapFnt{\small}
\SetAlCapNameFnt{\small}
\SetAlCapHSkip{0pt}

\definecolor{green}{rgb}{0, 0.5, 0}
\definecolor{orange}{rgb}{0.6, 0.3, 0.1}
\definecolor{red}{rgb}{1.0, 0.0, 0.0}
\definecolor{teal}{rgb}{0.0, 0.4, 0.4}
\definecolor{purple}{rgb}{0.65,0,0.65}
\definecolor{saffron}{rgb}{0.95,0.75,0.2}
\definecolor{turquoise}{rgb}{0.0,0.5,0.5}
\definecolor{brown}{rgb}{0.5, 0.16, 0.16}
\definecolor{brickred}{rgb}{.6, .2 .1}
\definecolor{coral}{rgb}{1,0.45,0.33}
\definecolor{newcolor}{rgb}{.8,.349,.1}

\definecolor{rankonecolor}{HTML}{DDF1D4}   
\definecolor{ranktwocolor}{HTML}{FFFAD4}   
\definecolor{rankthreecolor}{HTML}{EAF2FF} 

\newcommand{\rankone}{\cellcolor{rankonecolor}}
\newcommand{\ranktwo}{\cellcolor{ranktwocolor}}
\newcommand{\rankthree}{\cellcolor{rankthreecolor}}

\makeatletter
\patchcmd{\@maketitle}{\vspace*{24pt}}{\vspace*{10pt}}{}{}
\makeatother

\begin{document}

\twocolumn[
    \maketitle
    \vspace{-1.5em}
    \begin{center}
    \centering
    \includegraphics[width=0.99\linewidth]{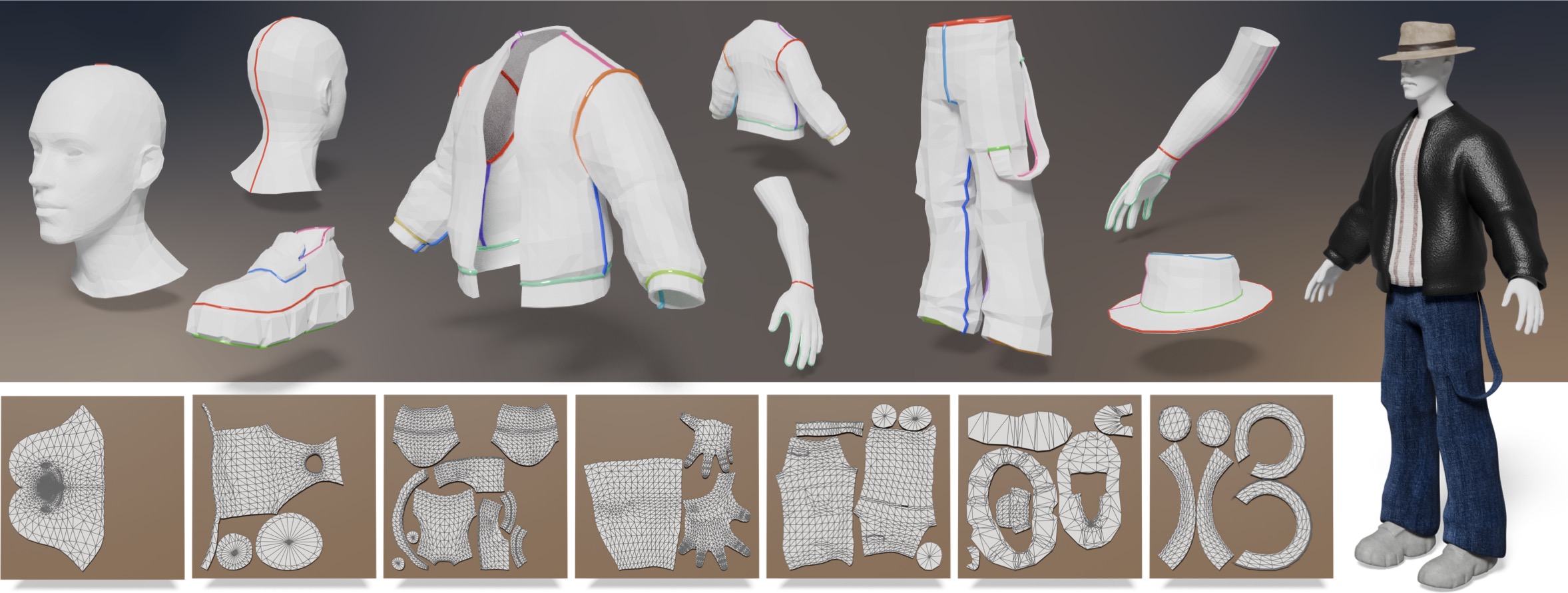}
    \captionof{figure}{
        \textbf{MeshTailor.}
        \textit{Top:} 
        MeshTailor generates seams (colored lines) directly on 3D meshes, producing clean, semantically aligned cuts that respect natural shape structure.
        \textit{Bottom:}
        The resulting seams partition surfaces into coherent UV charts, which are flattened into 2D layouts with minimal fragmentation.
        \textit{Right:} These high-quality UV maps facilitate seamless texture application, as demonstrated by the final textured character model.
    }
    \label{fig:teaser}
\end{center}

    \bigbreak
]


\let\thefootnote\relax\footnotetext{
$^*$Corresponding author.
}

\begin{abstract}
We present MeshTailor, the first mesh-native generative framework for synthesizing edge-aligned seams on 3D surfaces. 
Unlike prior optimization-based or extrinsic learning-based methods, MeshTailor operates directly on the mesh graph, eliminating projection artifacts and fragile snapping heuristics. 
We introduce ChainingSeams, a hierarchical serialization of the seam graph that orders chains from global structural cuts down to local details in a coarse-to-fine manner, and a dual-stream encoder that fuses topological and geometric context.
Leveraging this hierarchical representation and dual-stream vertex embeddings, our MeshTailor Transformer utilizes an autoregressive pointer layer to trace seams vertex-by-vertex within local neighborhoods. Extensive evaluations show that MeshTailor produces more coherent and structurally regular seam layouts compared to recent optimization-based and learning-based baselines.
\end{abstract}

\section{Introduction}

Flattening a 3D surface into 2D charts hinges on where to cut, and the best cuts follow the shape's own structural logic rather than local geometry alone.
Like a tailor placing seams along a body's contours, practitioners read a 3D shape's intrinsic geometry through curvature extrema and symmetries to trace long, coherent boundaries.
These boundaries carry design intent while respecting object form and function, and they underpin downstream applications including texturing, garment design, and digital fabrication.

However, relying on manual expertise creates a significant bottleneck in production pipelines, particularly for UV mapping in texture workflows. 
To automate this, classical~\cite{OptCuts} and neural-based~\cite{Nuvo,FAM} methods typically formulate seam generation as an optimization problem, balancing metrics such as distortion and cut length.
Classical formulations provide rigorous, data-free guarantees and remain the backbone of many production pipelines. What they leave open is the higher-level structural logic, namely the correspondence between shape morphology and cut placement that guides professionals, since purely local energies rarely encode it.
Consequently, boundaries derived from optimization may traverse natural symmetries or feature lines, yielding charts that, while low in distortion, may lack the semantic organization expected in artistic workflows.

Recent learning-based methods have opened a promising direction by distilling professional cutting strategies from data. However, representing precise seams on irregular meshes poses a fundamental challenge, and existing methods rely on extrinsic proxies that operate outside the native mesh graph. \textit{PartUV}~\cite{partuv}, for instance, predicts segmentation via volumetric fields. Because this proxy is decoupled from the surface geometry, mapping boundaries back to the mesh introduces misalignment with mesh edges, yielding irregular boundaries downstream.
More directly, SeamGPT~\cite{SeamGPT} proposes autoregressive seam generation by predicting vertex coordinates in Euclidean space. However, snapping predicted coordinates to mesh edges introduces projection artifacts and edge misalignment.

We focus on low-poly meshes (on the order of a few thousand triangles), the common asset regime in games and real-time rendering, where every edge already carries significant semantic weight. Working directly on the mesh would eliminate the projection artifacts above, yet mesh-native deep generation of seams remains largely unexplored in this regime.
This setting poses three core challenges:
First, meshes are irregular graphs lacking the uniform structure that enables efficient convolutions on images or grids.
Second, seams are not independent edge labels but form structured graphs with loops, chains, and hierarchical connectivity that need to be modeled explicitly.
Third, low-poly tessellation amplifies the cost of every decision: local geometric cues are sparse and noisy, chart boundaries are pinned to a small set of discrete edges with no room for smooth interpolation, and small placement errors translate directly into highly visible boundary defects. 
Faced with these difficulties, prior learning-based methods turn to extrinsic proxies, accepting the projection artifacts as the price of a tractable representation.

Drawing inspiration from the way tailors design seams to fit 3D forms, we introduce \textbf{MeshTailor}, a mesh-native generative framework for seams on low-poly meshes.
Unlike extrinsic methods that suffer from projection artifacts, MeshTailor operates natively on the mesh topology, reformulating seam placement as an autoregressive graph traversal via a \emph{Pointer Network}~\cite{PointerNet} over geometry-aware vertex embeddings. 
By restricting the decision space to the local 1-ring neighborhood, the model traces vertex sequences step-by-step. 
This approach guarantees mesh-aligned, continuous boundaries by construction, eliminating the projection step that introduces artifacts in extrinsic methods.

To enable this, we introduce \textbf{ChainingSeams}, a serialization scheme that imposes a hierarchical order on the seam chains by placing global structural cuts before local details.
We support this representation with a \textit{dual-stream encoder} that fuses topological graph features with geometric shape information represented as point clouds.
This hybrid embedding equips the pointer network with the rich context needed to trace these hierarchical paths accurately, yielding seam layouts that are both edge-aligned and semantically coherent. 
Extensive evaluations demonstrate that our approach produces seam layouts whose chart counts closely match artist-authored layouts, with substantially less fragmentation than greedy charting baselines and more regular, edge-aligned islands than optimization-based and recent neural baselines.

In summary, our key contributions are:

\begin{itemize} 
    \item \textbf{MeshTailor:} The first generative framework to synthesize seams intrinsically on the mesh, ensuring mesh-native, edge-aligned generation without extrinsic proxies or projection. 
    \item \textbf{ChainingSeams:} A hierarchical serialization scheme that organizes the seam graph from global partitioning cuts to local details, exposing a top-down structural progression to autoregressive decoders.
    \item \textbf{Efficient Graph Traversal:} A Pointer Network with a dual-stream (topological + geometric) encoder that restricts vertex selection to 1-ring neighborhoods, keeping the decision space compact and the autoregressive sequence short.
\end{itemize}

\begin{figure*}
    \centering
    \includegraphics[width=0.99\linewidth]{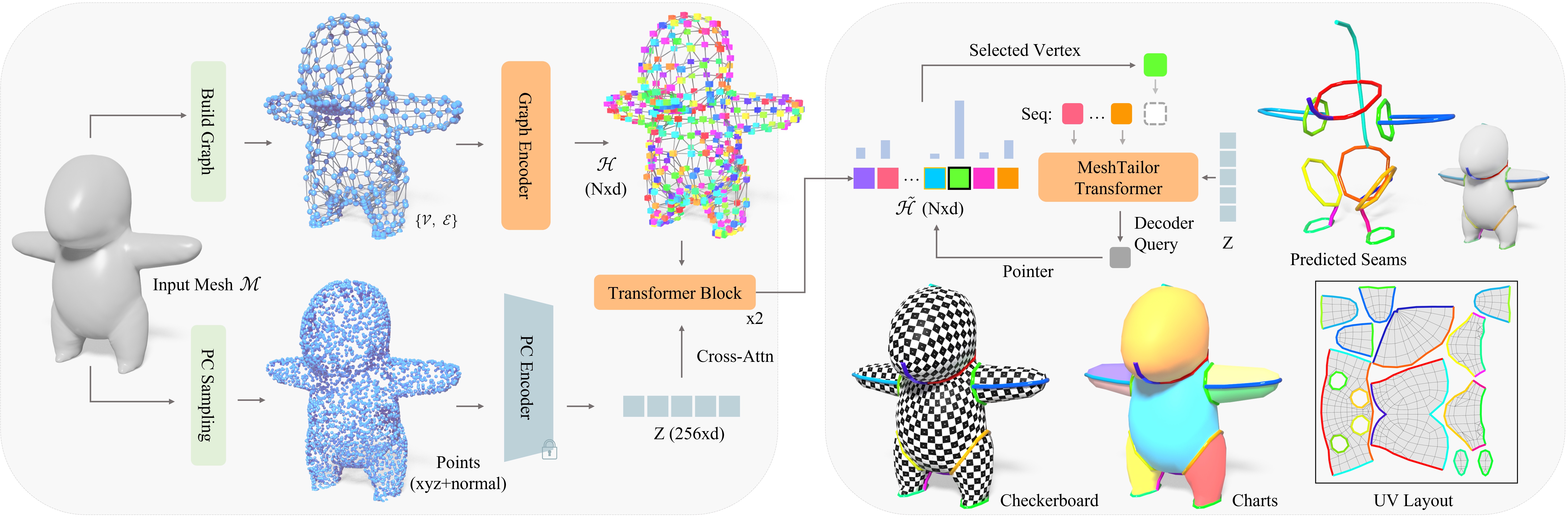}
    \caption{
\textbf{Overview of MeshTailor.}
    \textit{Left:} The dual-stream encoder.
    The input mesh is processed in parallel:
    the top stream extracts topological connectivity features $\mathcal{H}$ via a Graph Encoder on the mesh topology $\{V, E\}$,
    while the bottom stream samples surface points to extract global shape semantics tokens $Z$ using a pretrained point-cloud encoder (frozen during training).
    These representations are fused via cross-attention within Transformer Blocks.
    \textit{Right:} The autoregressive decoder.
    At each step, the MeshTailor Transformer conditions on the previously generated sequence (``Seq'') to produce a decoder query.
    A pointer layer attends to the enhanced vertex embeddings $\tilde{\mathcal{H}}$ to select the next vertex (green box), which is appended to the sequence.
    The resulting seam chains partition the mesh into UV charts, visualized here with checkerboard texturing to show low distortion, color-coded charts, and the final 2D UV layout.
    }
    \label{fig:overview}
\end{figure*}

\section{Related Work}
\label{sec:rw}
Seam placement for UV parameterization intersects several research areas, including classical surface flattening, mesh segmentation, and 3D generative modeling. In this section, we review prior methods in these areas, highlighting their limitations for production-ready low-poly assets, and motivate the need for mesh-native seam generation that preserves both geometric validity and stylistic consistency.

\subsection{Optimization-based surface parameterization}
Classical surface parameterization methods~\cite{Sheffer2006Survey,Hormann08} optimize geometric objectives under a fixed chart topology, providing rigorous distortion and bijectivity guarantees, and remaining the production backbone of UV mapping.
Foundational barycentric embeddings~\cite{tutte1963draw,Floater97,Eck95,FLOATER200319} place each interior vertex at a weighted average of its neighbors, while broader formulations minimize distortion via conformal~\cite{LSCM,Mullen08,BFF}, angle-based~\cite{ABF,ABFpp}, and locally/globally injective~\cite{SLIM,Du2021InjectiveParam} mappings.
Joint methods relax this assumption and couple cut placement with parameterization~\cite{Sorkine02,Autocuts,OptCuts,VariationalCuts}.

More recently, neural-based methods such as Nuvo~\cite{Nuvo} and FAM~\cite{FAM} learn UV maps end-to-end, bundling chart decomposition and parameterization within a per-shape neural optimization.
While they improve robustness on irregular 3D representations, they remain objective-driven: cuts emerge as byproducts of distortion minimization, leaving stylistic criteria, such as semantic alignment and coherent loops, largely unaddressed.
In contrast to these end-to-end pipelines, MeshTailor decouples seam generation from parameterization, learning seam priors that produce mesh-native cut layouts directly consumable by the classical solvers above that require a fixed chart topology.

\subsection{Mesh segmentation}
Mesh segmentation decomposes surfaces into meaningful regions whose boundaries become seams; classical methods use geometric cues (curvature, dihedral angles, spectral analysis) to form UV charts via face clustering~\cite{Sander2001TMPM}, quasi-developable segmentation~\cite{DCharts}, and spectral stretch-driven charting~\cite{Isocharts}.
Production toolchains such as xatlas~\cite{xatlas_github} and Blender~\cite{blender} apply similar greedy splitting heuristics; while robust across diverse assets, they offer no explicit stylistic control and often over-segment, producing excessive fragmentation.
Learning-based segmentation methods inject data-driven priors. Neural part segmentation~\cite{PartField,yang2024sampart3d} provides general shape decompositions but without explicit UV objectives. PartUV~\cite{partuv} builds on learned part trees for top-down recursive unwrapping, while GraphSeam~\cite{GraphSeam} learns a discriminative per-edge seam classifier with GraphSAGE~\cite{SAGEConv} and refines its outputs via postprocessing. DA-Wand learns distortion-aware patch selection, treating seams implicitly as patch boundaries~\cite{DAWand}.
Unlike these pipelines, which treat seams as \emph{byproducts of decomposition} or \emph{independent edge labels} and leave structure to heuristics or postprocessing, MeshTailor directly models seam structure as ordered vertex walks, yielding well-formed seam graphs (closed loops and complete chains) without dangling edges or postprocessing.

\subsection{Generative 3D modeling}
Seam placement can also be considered as a 3D generation problem.
Volumetric representations~\cite{xie2022neural} are commonly used for 3D generation via diffusion~\cite{3DShape2VecSet, xiang2024structured} or autoregressive models~\cite{ShapeFormer, mittal2022autosdf, wei2025octgpt}, but they lose the underlying surface geometry and connectivity.
Mesh-based generative models preserve such information by learning directly on mesh data, generating new meshes as polygonal graphs~\cite{PolyGen}, text~\cite{llamamesh}, triangle sequences~\cite{MeshGPT,Meshtron}, or UV charts~\cite{omages,GIMDiffusion,li2025garmagenet}.

Closer to our work, structural-subgraph generators such as wireframe models~\cite{WireframeGen} produce ordered vertex--edge graphs. Seam generation is a special case: SeamGPT~\cite{SeamGPT} autoregressively predicts seam-vertex coordinates that are then snapped to mesh edges, and SeamCrafter~\cite{SeamCrafter} refines seam quality via reinforcement learning.
However, these methods either generate new wireframes, or predict 3D seam-vertex coordinates that are then snapped to mesh edges. In contrast, MeshTailor traverses the given mesh's existing vertices and edges, producing seam subgraphs as ordered vertex indices, without snapping or new geometry.

\begin{figure*}
    \centering
    \includegraphics[width=0.99\linewidth]{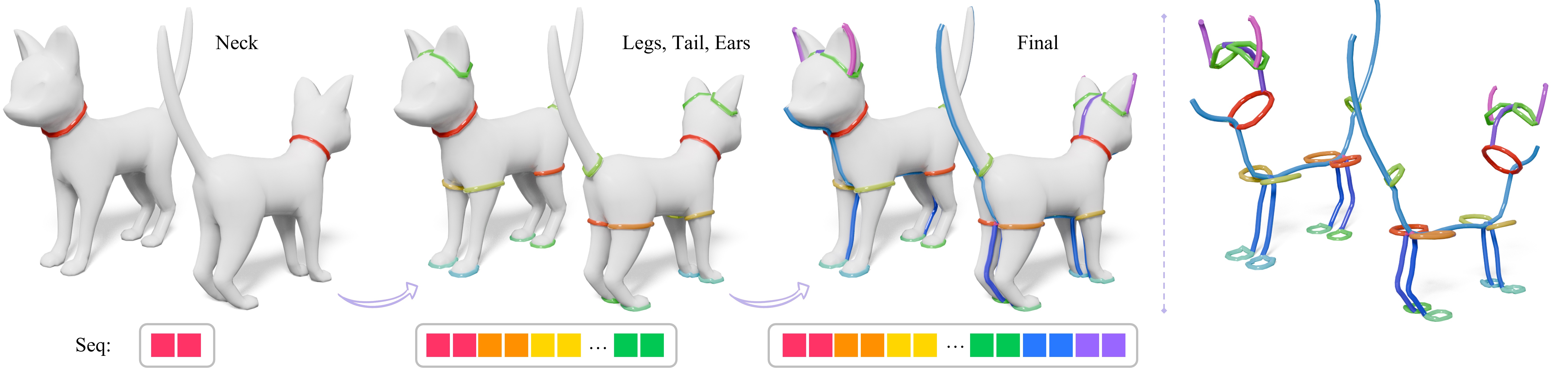}
    \caption{
    \textbf{Canonical ordering of seam chains (coarse-to-fine).}
    We serialize an unordered seam set into a deterministic sequence for autoregressive training/inference with a \emph{loops-first, balance-first, large-patch-first} strategy: we prioritize loop cuts over open chains, repeatedly select the \emph{largest} remaining surface patch, and within that patch choose the loop cut that best \emph{balances} the two resulting sub-patch areas (see Supplemental Material~\ref{sec:supp_ordering} for details).
    The example shows how primary loop cuts (e.g., around the neck) are placed first to decompose the shape, followed by finer loops on smaller parts (legs, tail, ears), yielding the final ordered chain list and token sequence $\boldsymbol{\tau}$.
    }    
    \label{fig:chains_order}
\end{figure*}

\begin{figure}[t]
  \centering
  \includegraphics[width=\linewidth]{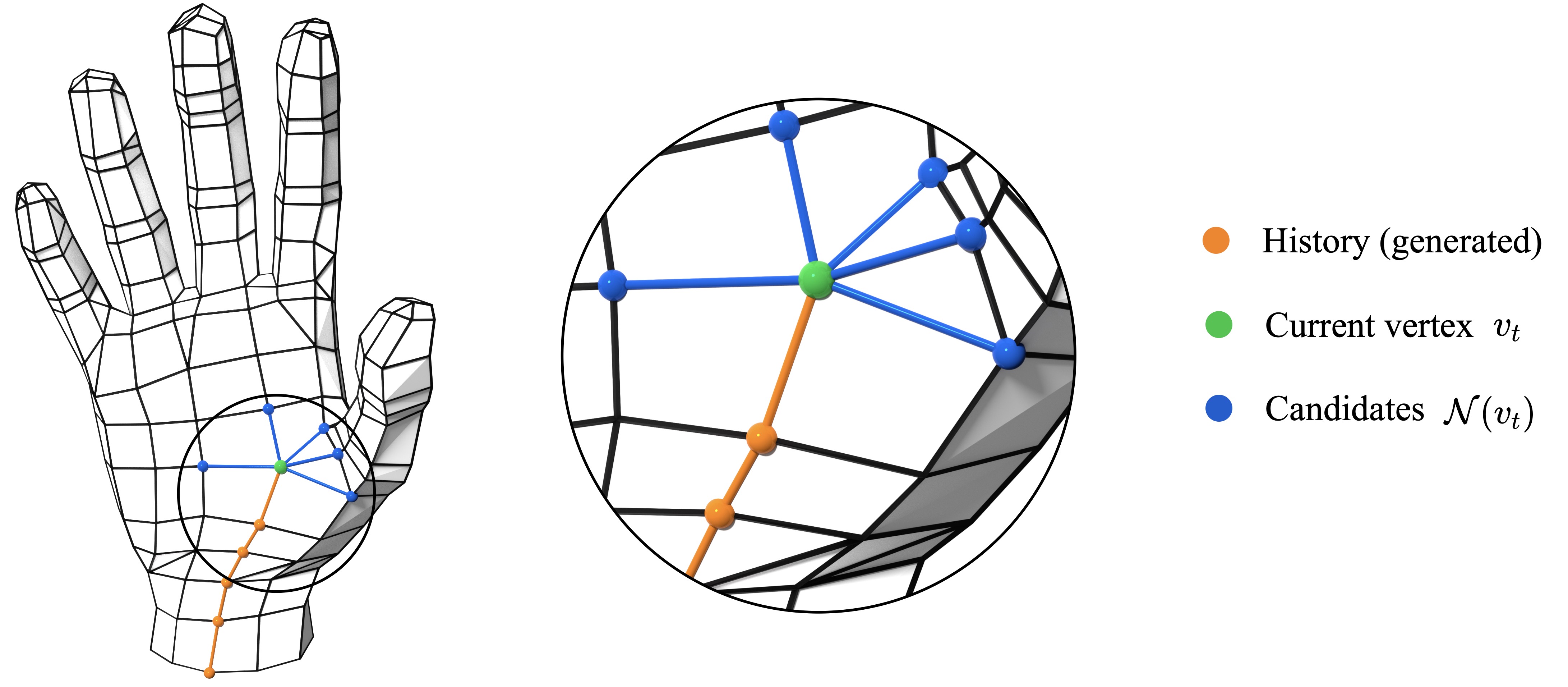}
  \caption{
  \textbf{Step-by-step seam generation as mesh traversal.}
    Our autoregressive decoder traverses the input mesh connectivity, selecting the next vertex from the current 1-ring neighborhood. 
    This local constraint makes the predicted seams mesh-native and avoids invalid jumps across the surface.
  }
  \vspace{-3mm}
  \label{fig:traversal_step}
\end{figure}

\section{Method}
\label{sec:method}

\subsection{Overview}
\label{sec:overview}

Our goal is to generate UV seams that are both \emph{edge-aligned and continuous} on the input mesh and \emph{semantically coherent}. We cast this task as an \emph{autoregressive graph traversal} on the input mesh $\mathcal{M}=(\mathcal{V},\mathcal{E},\mathcal{F})$. \textbf{MeshTailor} predicts a collection of \emph{seam chains} $\mathcal{C}$, where each chain is a sequence of connected edges in $\mathcal{E}$ represented as a vertex walk $\mathbf{c}=(v_1,\dots,v_T)$. Since a mesh typically requires multiple such chains, the seam layout forms an unordered graph, whereas autoregressive modeling requires a sequentialized representation. To address this, we introduce \emph{ChainingSeams}~(Sec.~\ref{sec:chainingseams}): a unified representation that decomposes the seam graph into seam chains and serializes them using a stable, coarse-to-fine ordering.

As illustrated in Fig.~\ref{fig:overview}, the framework operates in two stages on top of our \emph{ChainingSeams} representation (Sec.~\ref{sec:chainingseams}). First, a \emph{dual-stream encoder} (Sec.~\ref{sec:dualstream}) fuses local mesh connectivity with global shape semantics to produce context-aware embeddings for every vertex. Second, the MeshTailor Transformer (Sec.~\ref{sec:meshtailor}) utilizes a pointer layer to autoregressively traverse the graph guided by these embeddings to generate edge-aligned seam chains. We train end-to-end with autoregressive next-token prediction over the serialized seam chains (Sec.~\ref{sec:traininginference}).

\subsection{ChainingSeams}
\label{sec:chainingseams}

ChainingSeams represents seams as ordered \emph{seam chains} rather than as independent edge labels. A seam chain $\mathbf{c} = (v_1, \dots, v_T)$ is a vertex walk on the mesh where every step corresponds to a mesh edge, i.e., $(v_t, v_{t+1}) \in \mathcal{E}$. This representation naturally supports both open chains and closed loops ($v_1 = v_T$), and exposes edge-alignment as a structural property that the decoder can later enforce (Sec.~\ref{sec:meshtailor}). Supplemental Material~\ref{sec:supp_representation} formalizes this representation and details the chain extraction procedure.

\paragraph{Canonical Ordering.}
We define a deterministic ordering with three principles, applied in priority order: (i) \emph{loops first} (closed-loop cuts before open chains), (ii) \emph{largest patch first} (among loop cuts, target the largest current surface patch), and (iii) \emph{area balance} (among such cuts, split the patch into sub-patches of roughly equal area). Remaining open chains are appended at the end. This yields a stable, coarse-to-fine ordering (see Fig.~\ref{fig:chains_order}) that mirrors artist UV-marking conventions~\cite{unwrella2009uvtutorial}, which progressively separate coarse parts before refining details; the full splitting algorithm is provided in Supplemental Material~\ref{sec:supp_ordering}.

\subsection{Dual Stream Encoder}
\label{sec:dualstream}

Producing seams that are both \emph{edge-aligned} and \emph{semantically coherent} requires two complementary signals: \emph{mesh connectivity} to route cuts along edges, and \emph{global shape semantics} to align cuts with object parts. We capture them with a dual-stream encoder.

First, to capture mesh-dependent connectivity, we treat the mesh $\mathcal{M}$ as a graph and process it with a GraphSAGE encoder~\cite{SAGEConv}, denoted $\mathrm{Enc}_G$, yielding per-vertex connectivity embeddings $\{\mathbf{h}_i\}_{v_i\in\mathcal{V}}$ with $\mathbf{h}_i \in \mathbb{R}^d$.
We stack them as $\mathcal{H}=[\mathbf{h}_i]_{i=1}^{N}$, where $\mathcal{H}\in\mathbb{R}^{N\times d}$.
Second, to capture tessellation-invariant shape semantics, we sample a surface point cloud and extract shape tokens $\mathbf{Z} \in \mathbb{R}^{M \times d}$ using a frozen pretrained point cloud encoder~\cite{Michelangelo}, denoted $\mathrm{Enc}_P$. Finally, we design a cross-attention fusion that injects global shape information into the local vertex representations:
\begin{equation}
\tilde{\mathbf{h}}_i = \mathrm{CrossAttn}(\text{query}=\mathbf{h}_i, \text{key/value}=\mathbf{Z}),
\end{equation}
producing enhanced embeddings $\tilde{\mathbf{h}}_i \in \mathbb{R}^d$ passed to the decoder (Sec.~\ref{sec:meshtailor}),
which we stack as $\tilde{\mathcal{H}}=[\tilde{\mathbf{h}}_i]_{i=1}^{N}\in\mathbb{R}^{N\times d}$.
Architecture details, including Fourier coordinate features and tokenization, are provided in Supplemental Material~\ref{sec:supp_arch}.

\subsection{MeshTailor Transformer}
\label{sec:meshtailor}

We generate seam chains autoregressively. Standard Transformer decoders predict over a fixed vocabulary, but our output space (mesh vertices) varies per input mesh. We therefore design a \emph{mesh-native pointer layer} that selects the next vertex directly from the input mesh. At step $t$, the decoder hidden state $\mathbf{q}_t \in \mathbb{R}^d$ queries the candidate embeddings to compute a probability distribution over candidates (mesh vertices and the control tokens \texttt{[EOC]}, \texttt{[EOS]}):
\begin{equation}
p(\tau_{t+1}=u \mid \tau_{\le t}) = \mathrm{softmax}_u\big( \langle \mathbf{q}_t, \mathbf{W}\,\mathbf{e}_u \rangle + m_{t,u} \big),
\label{eq:pointer}
\end{equation}
where $\mathbf{e}_u$ is the embedding of candidate $u$ (the per-vertex embedding $\tilde{\mathbf{h}}_u$ for a vertex, or a learned token embedding for \texttt{[EOC]}/\texttt{[EOS]}) and $\mathbf{W} \in \mathbb{R}^{d \times d}$ is a learned matrix.
We enforce edge-alignment via the mask term $m_{t,u}$. We set $m_{t,u}=0$ if candidate $u$ is a 1-ring neighbor of the current vertex $v_t$ or a control token (i.e., \texttt{[EOC]} or \texttt{[EOS]}), and $m_{t,u}=-\infty$ otherwise. This constrains generation to vertex walks along mesh edges, ensuring each chain is edge-aligned by construction without requiring post-hoc projection. Each token additionally receives a chain-local positional embedding that resets after every \texttt{[EOC]} to encode within-chain progress (Sec.~\ref{sec:supp_arch}). Fig.~\ref{fig:traversal_step} illustrates this mesh-native graph traversal.

\subsection{Training and Inference}
\label{sec:traininginference}

\paragraph{Training.}
We supervise MeshTailor with seam annotations extracted from UV-annotated datasets (Sec.~\ref{sec:impl}). For each mesh, we convert these unordered seam edges into a deterministic target sequence $\boldsymbol{\tau}$ using the canonical ordering strategy described in Sec.~\ref{sec:chainingseams}. The sequence concatenates the vertex indices of all chains, with consecutive chains separated by an \texttt{[EOC]} token and the full sequence terminated by an \texttt{[EOS]} token. 
We train the model to minimize the negative log-likelihood of the next token:
\begin{equation}
\mathcal{L}_{\mathrm{AR}} = - \sum_{t} \log p(\tau_{t+1} \mid \tau_{\le t}, \mathbf{Z}),
\label{eq:ar_obj}
\end{equation}
where $\mathbf{Z}$ represents the global shape context.

\paragraph{Inference.}
During inference, we compute the encoder embeddings and generate seams autoregressively. 
The model samples a start vertex from $\mathcal{V}$ (without the neighbor mask), then autoregressively walks along mesh edges under the mask, excluding immediate backtracking, until predicting \texttt{[EOC]}.
This process repeats until \texttt{[EOS]} is generated, yielding the final seam layout $\mathcal{C}$.
The inference pseudocode is provided in Supplemental Material~\ref{sec:supp_infer_pseudocode}.

\begin{figure*}
    \centering
    \includegraphics[width=0.95\linewidth]{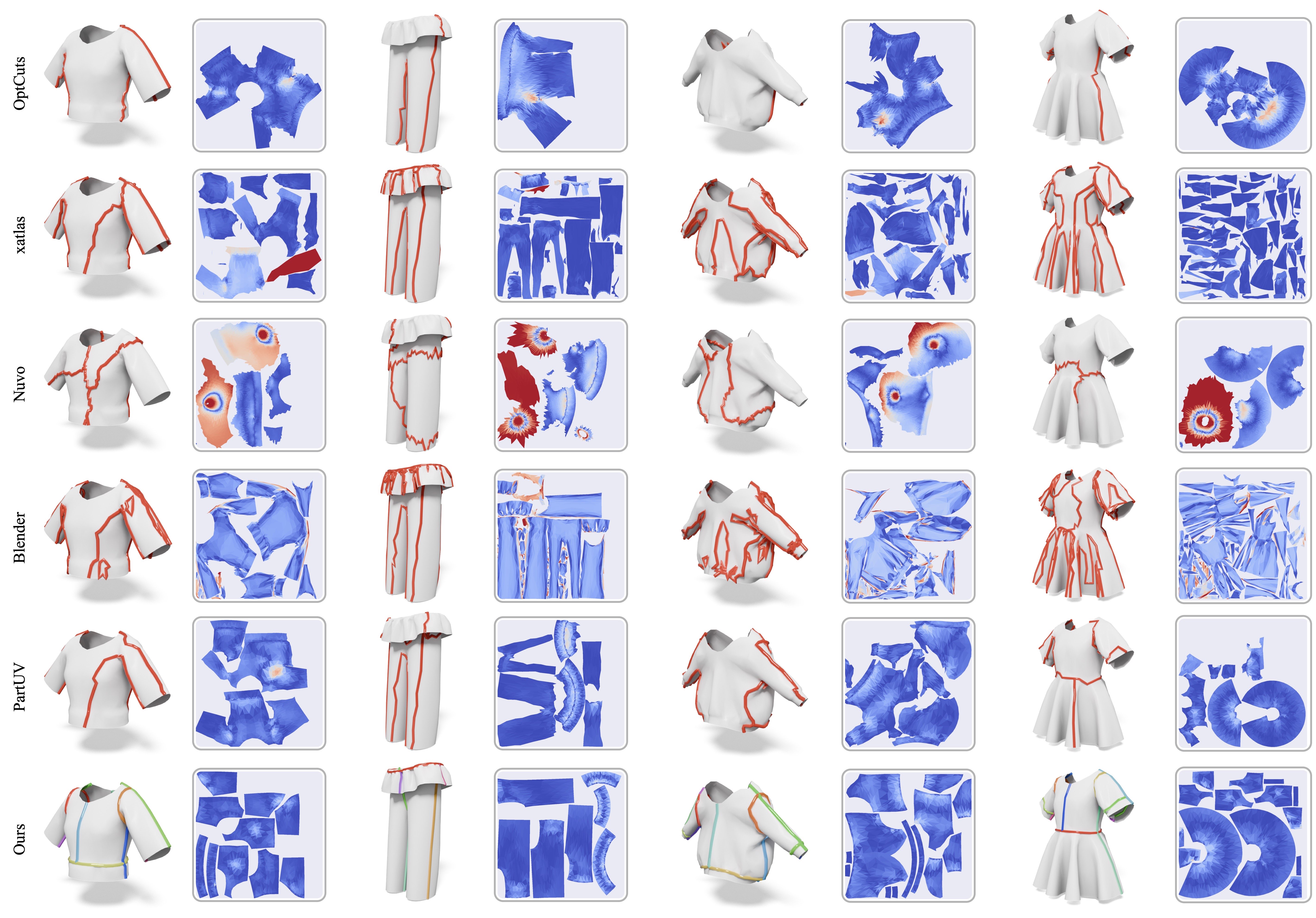}
    \caption{
    \textbf{Seam layout and area distortion comparison on GarmentCodeData \cite{GarmentCodeData}.}
    For each method, we show the predicted seams on the 3D mesh (left) and the corresponding area distortion heatmap on the UV layout (right).
    While prior methods often produce fragmented or jagged cuts that lead to irregular UV islands, MeshTailor generates cleaner, garment-aligned seam structures with coherent chains and loops, resulting in more regular, compact UV charts while maintaining competitive distortion.
    Different colors in our seam visualizations indicate distinct seam chains (and loop cuts), highlighting the structured and editable output of our representation.
    }
    \label{fig:garment_seam}
\end{figure*}

\section{Experimental Results}
\label{sec:results}

\subsection{Implementation Details}
\label{sec:impl}

\paragraph{Datasets and preprocessing.}
We train and evaluate on two large-scale datasets: TexVerse~\cite{TexVerse} and GarmentCodeData~\cite{GarmentCodeData}. 
After filtering for high-quality UV layouts, decomposing into part-level components, and removing trivial shapes, we obtain 300K part-level and 110K garment samples. 
We split each dataset into 90\%/5\%/5\% for train/validation/test.
See Supplemental Material~\ref{sec:supp_data} for complete preprocessing details.

\paragraph{Supervision and training.}
We extract seam edges from UV layouts using Blender's seam extraction operator~\cite{blender_seams_from_islands}, trace them into chains, and apply canonical ordering (Supplemental Material Alg.~\ref{alg:chain_order}).
We train with AdamW (learning rate $10^{-4}$, batch size 64) for 30 epochs on 8 NVIDIA RTX 4090 GPUs, sampling 2,048 surface points per mesh.
Since we decode seams as vertex indices rather than 3D coordinates, the autoregressive sequences are considerably shorter in practice; we cap the decoding length at $T_{\max}=400$ in all experiments.
The point-cloud encoder is pretrained and frozen; we train the graph encoder and MeshTailor Transformer end-to-end.

\subsection{Evaluation Protocol and Metrics}
\label{sec:metrics}

After seam generation, we flatten each chart using ABF++~\cite{ABFpp}.
We evaluate along four complementary axes:
(i) \textit{Distortion}, following PartUV~\cite{partuv}, including overall (area) distortion aggregated from triangle-wise area stretch over the atlas, and angle preservation (higher is more conformal);
(ii) \textit{UV island quality}, measured by compactness $4\pi\,\mathrm{area}/\mathrm{perimeter}^2$ and convexity $\mathrm{area}/\mathrm{area}(\mathrm{convex\ hull})$;
(iii) \textit{Seam properties}, including normalized seam length (per 3D area) and boundary jaggedness of UV chart outlines, computed via a discrete curvature proxy on resampled UV boundary loops (lower is smoother);
and (iv) \textit{Structural complexity}, measured by the number of charts (UV islands).
Full definitions, including the discrete curvature proxy for jaggedness, are provided in Supplemental Material~\ref{sec:supp_metrics}.

\subsection{Comparison with Baselines}
\label{sec:comparison}

\paragraph{Baselines.}
We compare against representative methods spanning several widely used pipelines in production toolchains, optimization-based cutting, and learned parameterization:
xatlas \cite{xatlas_github} (greedy charting with LSCM flattening),
Blender Smart UV Project \cite{blender} (bottom-up chart clustering guided by local normals),
OptCuts \cite{OptCuts} (joint cut-and-distortion optimization under global constraints),
Nuvo \cite{Nuvo} (learned continuous UV field), and
PartUV \cite{partuv} (part-aware UV pipeline with reduced chart counts).

\begin{table}[t]
\centering
\caption{
    \textbf{Quantitative comparison on TexVerse and GarmentCodeData.}
    Note that Nuvo is evaluated on 100 randomly sampled meshes due to runtime, while all other methods use the full test set.
    We additionally report GT on both datasets, where GT denotes seams extracted from the dataset UV layouts and is shown only as an extra reference.
    For each metric, the top-3 results among the compared methods are highlighted as \colorbox{rankonecolor}{best}, \colorbox{ranktwocolor}{second}, and \colorbox{rankthreecolor}{third}.
}
\label{tab:main}
\scriptsize
\setlength{\tabcolsep}{3.0pt}
\renewcommand{\arraystretch}{1.12}

\begin{tabular}{lccccccc}
\toprule
\multicolumn{8}{c}{\textbf{TexVerse}} \\
\midrule
Method
& \shortstack{Overall\\Dist$\downarrow$}
& \shortstack{Angular\\Dist$\uparrow$}
& \shortstack{\#\\Charts$\downarrow$}
& \shortstack{Island\\Compact$\uparrow$}
& \shortstack{Island\\Convex$\uparrow$}
& \shortstack{SeamLen\\/Area$\downarrow$}
& \shortstack{Boundary\\Jagged$\downarrow$} \\

\midrule
Blender                 & 1.182 & 0.753 & 48.893 & 0.287 & \rankthree0.753 & 6.356 & 1.313 \\
xatlas                  & \rankone1.087 & \rankone0.975 & 47.100 & \rankthree0.412 & 0.752 & 6.785 & \rankthree1.092 \\
PartUV                  & \ranktwo1.134 & \ranktwo0.953 & 16.248 & 0.385 & 0.742 & \rankthree4.195 & \ranktwo1.061 \\
OptCuts                 & \rankthree1.139 & 0.887 & \rankone1.6 & 0.309 & 0.739 & \rankone2.190 & 1.192 \\
$\mathrm{Nuvo}^*$       & 4.749 & 0.846 & \ranktwo6.42 & \rankone0.583 & \rankone0.854 & 5.628 & 1.448 \\
\midrule
Ours                    & 1.218 & \rankthree0.927 & \rankthree8.631 & \ranktwo0.541 & \ranktwo0.840 & \ranktwo2.784 & \rankone0.686 \\
\midrule
GT                      & 1.193 & 0.945 & 9.021 & 0.572 & 0.847 & 2.837 & 0.641 \\
\midrule
\multicolumn{8}{c}{\textbf{GarmentCodeData}} \\
\midrule
Method
& \shortstack{Overall\\Dist$\downarrow$}
& \shortstack{Angular\\Dist$\uparrow$}
& \shortstack{\#\\Charts$\downarrow$}
& \shortstack{Island\\Compact$\uparrow$}
& \shortstack{Island\\Convex$\uparrow$}
& \shortstack{SeamLen\\/Area$\downarrow$}
& \shortstack{Boundary\\Jagged$\downarrow$} \\

\midrule
Blender              & 1.147 & 0.938 & 74.265 & 0.203 & \ranktwo0.820 & 6.513 & 1.440 \\
xatlas               & \rankone1.064 & \rankone0.974 & 51.648 & \rankthree0.358 & 0.783 & 5.930 & 1.188 \\
PartUV               & 1.113 & \rankthree0.969 & \ranktwo5.198 & \ranktwo0.438 & 0.789 & \ranktwo1.730 & \ranktwo0.882 \\
OptCuts              & \ranktwo1.082 & 0.937 & \rankone1.09 & 0.328 & 0.763 & \rankone0.729 & \rankthree1.050 \\
$\mathrm{Nuvo}^*$    & 2.366 & 0.956 & \rankthree6.49 & 0.352 & \rankthree0.803 & 6.423 & 2.310 \\
\midrule
Ours                 & \rankthree1.097 & \ranktwo0.970 & 10.414 & \rankone0.591 & \rankone0.887 & \rankthree2.25 & \rankone0.485 \\
\midrule
GT                   & 1.095 & 0.972 & 11.449 & 0.592 & 0.891 & 2.398 & 0.473 \\
\bottomrule
\end{tabular}

\vspace{-4mm}
\end{table}

\paragraph{Quantitative results}
\label{sec:quant}

Table~\ref{tab:main} summarizes results on TexVerse and GarmentCodeData.
While xatlas achieves the lowest distortion, it produces highly fragmented UV layouts with many small charts.
Compared to Blender and xatlas, our method produces substantially fewer charts while improving island regularity (compactness/convexity) and reducing seam jaggedness, at the cost of slightly higher but competitive distortion.
Compared to PartUV, our method yields seam layouts that are more mesh-aligned and structurally coherent (smoother boundaries), resulting in islands that are easier to edit in DCC tools.
Optimization-based baselines (OptCuts, Nuvo) can reduce chart counts, but may still generate seam boundaries that are not aligned with production seam conventions (e.g., overly irregular or difficult-to-edit boundaries).

\begin{figure*}
    \centering
    \includegraphics[width=0.97\linewidth]{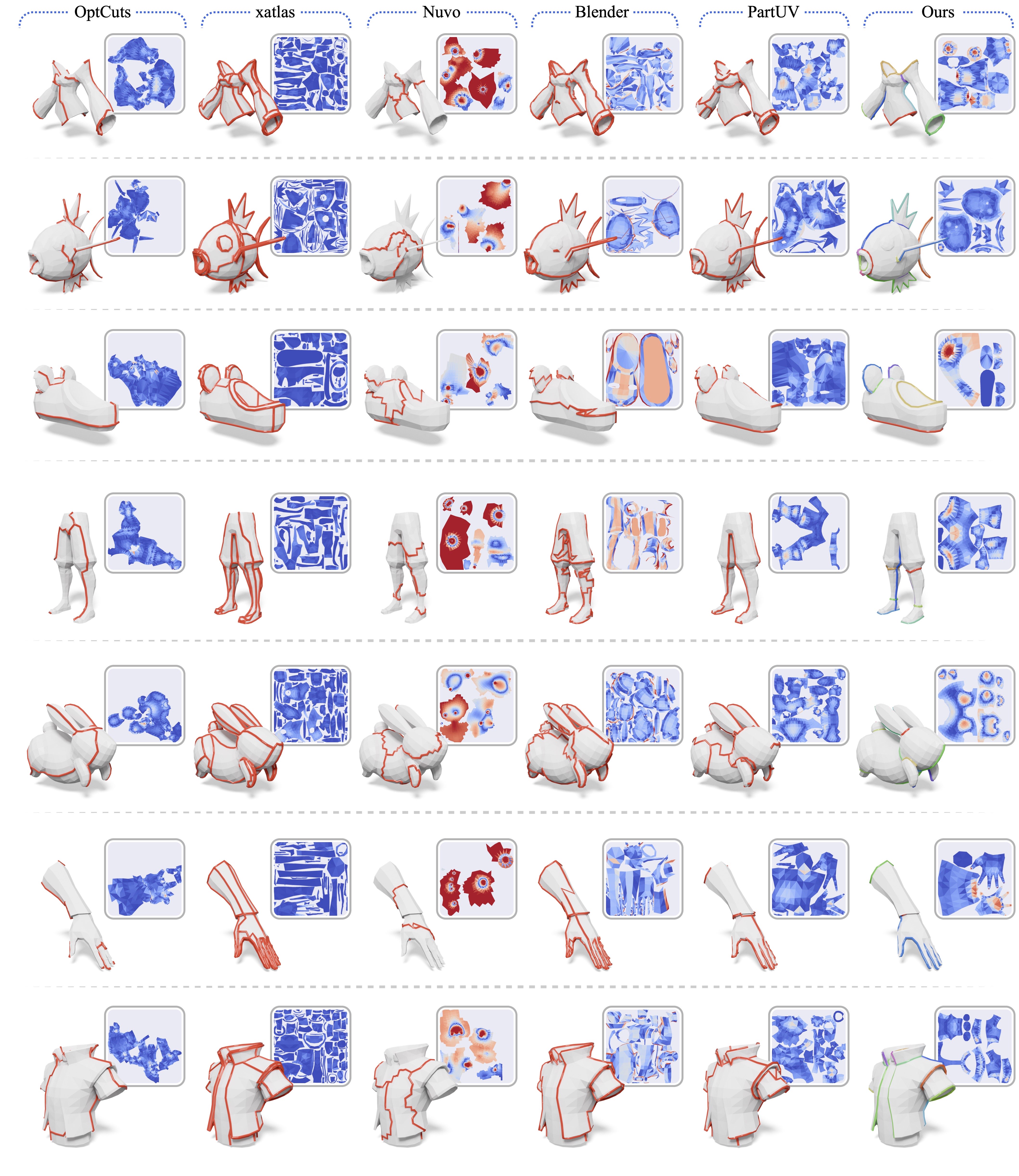}
    \caption{
    \textbf{Seam layout and UV area distortion comparison on TexVerse}~\cite{TexVerse}.
    We qualitatively compare MeshTailor with OptCuts, xatlas, Nuvo, Blender Smart UV Project, and PartUV on diverse assets from TexVerse.
    Compared to existing baselines that often produce fragmented charts or jagged cut boundaries, MeshTailor generates cleaner, mesh-aligned seams with coherent long-range chain/loop structures while maintaining competitive distortion.
    Different colors in our results denote distinct seam chains/loops, highlighting the structured and editable seam representation.
    }
    \label{fig:texverse_seam}
\end{figure*}

\paragraph{Qualitative comparison}
\label{sec:qual}

Fig.~\ref{fig:garment_seam} and Fig.~\ref{fig:texverse_seam} compare predicted seam layouts and the corresponding UV area distortion heatmaps across methods.
Blender and xatlas tend to introduce many small charts with jagged boundaries, which complicates texture painting and increases visible discontinuities.
PartUV reduces chart counts by leveraging part priors, yet its seam boundaries can remain irregular, limiting downstream editing.
OptCuts produces extremely few charts with short seams, but the resulting islands can be less regular and its boundaries remain less clean than ours.
Nuvo often yields less stable seam layouts with higher distortion and boundary irregularities.
In contrast, our method produces mesh-aligned seam chains that form coherent long-range structures (including loop cuts), leading to cleaner, more regular UV islands with smoother, more editable boundaries while maintaining competitive distortion.

\begin{figure}[t]
  \centering
  \includegraphics[width=\linewidth]{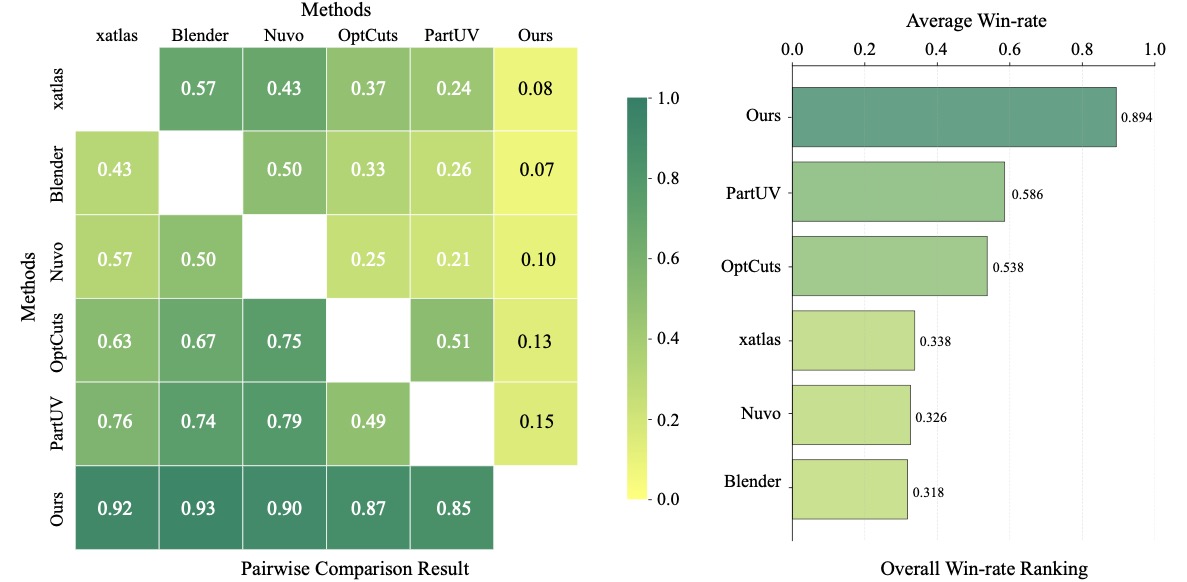}
  \caption{
    Left: pairwise preference rates between methods, where each entry reports the fraction of participants favoring the row method over the column method.
    Right: overall ranking by average win-rate against all other methods.
    MeshTailor is consistently preferred over all baselines.
  }
  \label{fig:user_study}
\end{figure}

\paragraph{User study.}
We conducted a 2AFC user study with 100 participants comparing seam layouts on three production criteria: minimization (fewer visible cuts), concealment (cuts in less salient regions), and geometry awareness (seams following natural geometric boundaries). In total, participants cast 5,000 pairwise votes.
We summarize the results as a pairwise preference matrix (Fig.~\ref{fig:user_study}).
For each ordered pair $(A,B)$, we report the preference rate $p(A\succ B)=n_A/(n_A+n_B)$, where $n_A$ and $n_B$ are the votes for $A$ and $B$ in head-to-head comparisons. $p(A\succ B)>0.5$ indicates that $A$ is preferred.
MeshTailor is consistently preferred over all baselines, supporting that mesh-native seam chains better match production conventions.
See Supplemental Material~\ref{sec:supp_user_study} for methodology details.

\begin{figure}[t]
  \centering
  \includegraphics[width=0.9\linewidth]{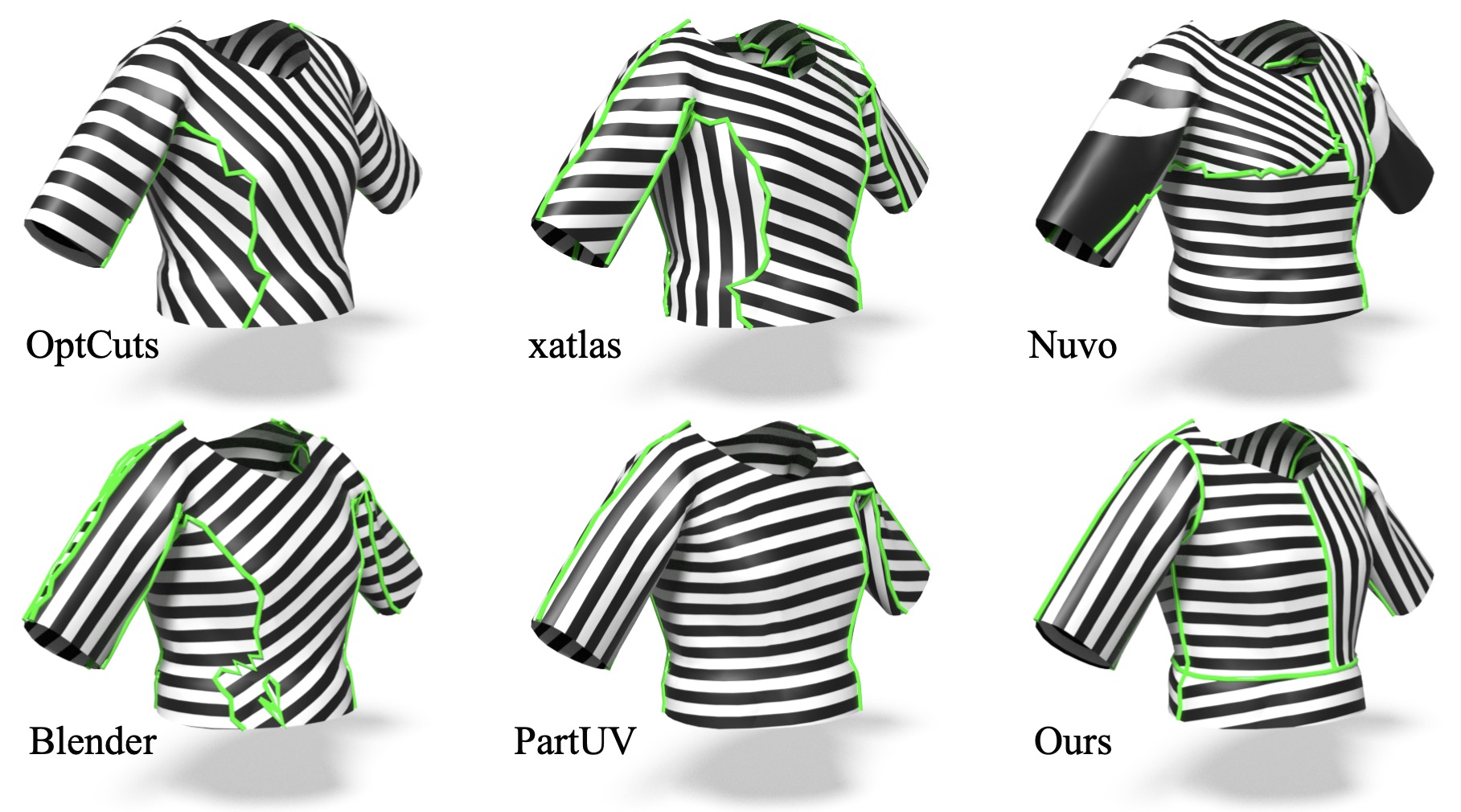}
  \caption{
  UV quality under a tiling stripe texture. 
  Fragmented charts and irregular boundaries often cause stripe discontinuities and misalignment, while MeshTailor preserves coherent regions with consistent texture flow.
  }
  \label{fig:uv_map}
\end{figure}

\begin{figure}[t]
  \centering
  \includegraphics[width=\linewidth]{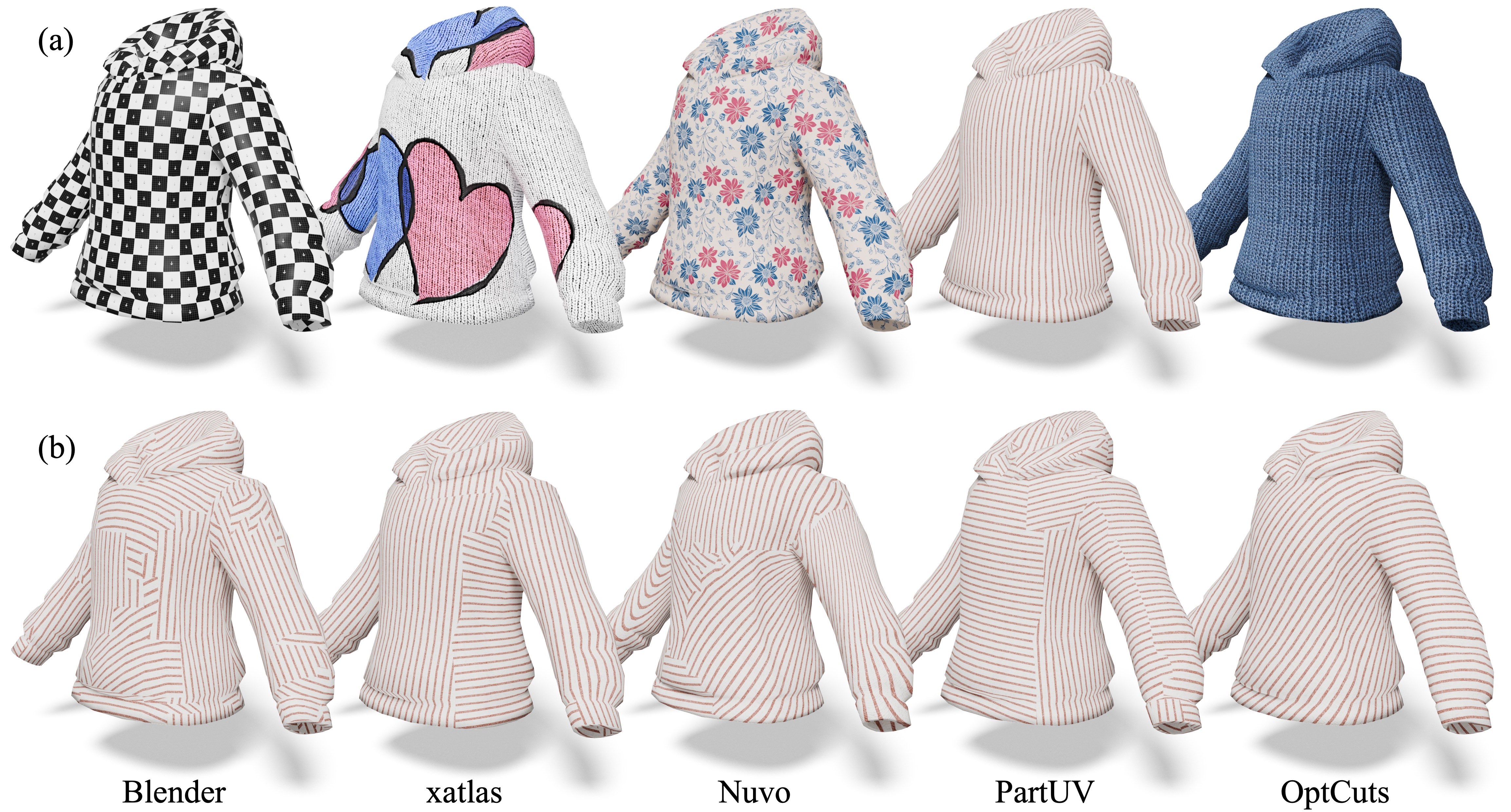}
  \caption{
    \textbf{Production-oriented UV usability.}
    (a) Our UV layout supports diverse texture appearances on the same garment.
    (b) Stripe tiling, showing results from prior methods.
  }
  \label{fig:texturing}
\end{figure}

\subsection{Applications}
\label{sec:apps}
\paragraph{Texture editing and replacement.}
Our coherent chart layout enables flexible texture editing and seamless tiled-texture application (Fig.~\ref{fig:texturing}).
Fig.~\ref{fig:texturing}(a) showcases \emph{MeshTailor} on the same garment with diverse texture applications (e.g., checkerboard, stylized patterns, and fabric-like materials), illustrating that our layout preserves coherent texture flow over large regions.
Fig.~\ref{fig:texturing}(b) compares methods under a tiling stripe texture: baseline methods with fragmented charts produce stripe discontinuities and boundary misalignment, while MeshTailor's coherent regions and smoother boundaries preserve stripes across the surface.

\begin{figure}[t]
  \centering
  \includegraphics[width=\linewidth]{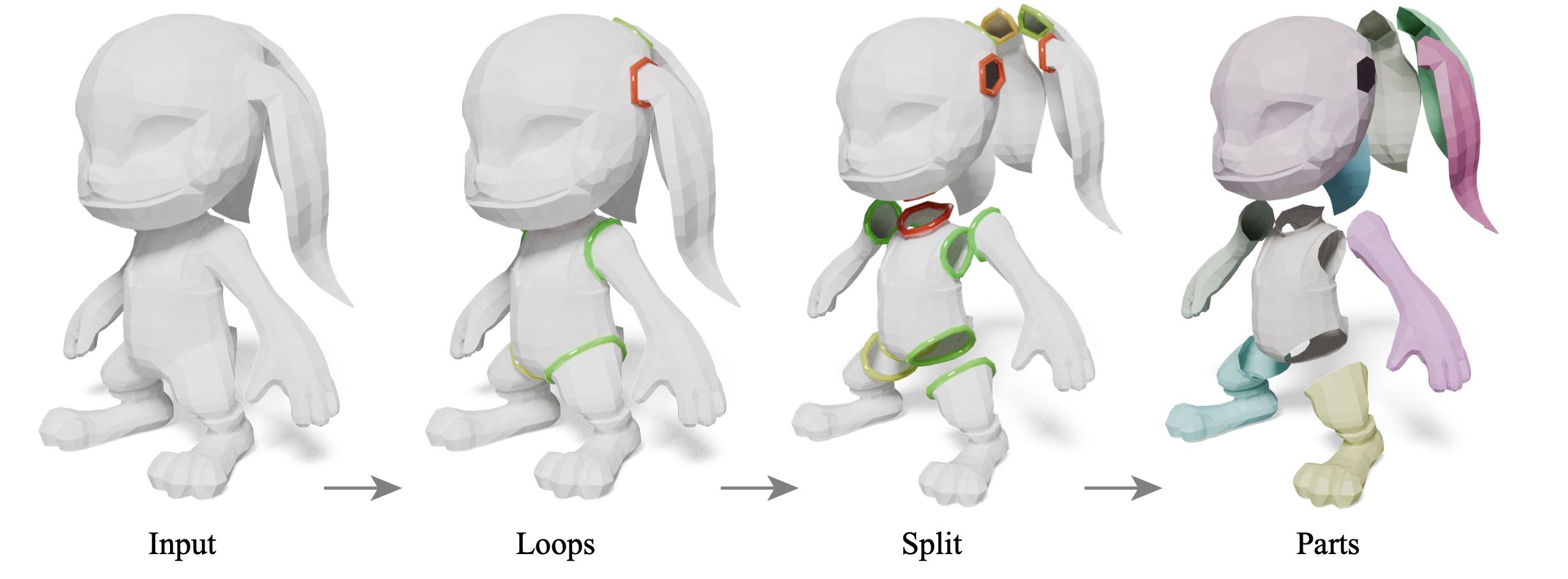}
  \caption{
    MeshTailor naturally produces loop cuts as part of its seam structure.
    Applying the predicted loops (middle), we recursively split the input mesh into connected patches with clean, artist-like boundaries, yielding a part decomposition (right) that is directly usable for downstream asset processing.
  }  
  \label{fig:part_seg}
\end{figure}

\paragraph{Loop-induced part segmentation.}
Loop cuts in our seam structure double as part boundaries, giving a mesh-aligned mechanism for coarse-to-fine surface decomposition.
By prioritizing high-level loops first, our method yields meaningful part boundaries (e.g., limb/torso separations on animals or garment components), supporting downstream part-aware processing (Fig.~\ref{fig:part_seg}).

\subsection{Robustness and Practical Considerations}
\label{sec:robust}
\begin{figure}[t]
  \centering
  \includegraphics[width=\linewidth]{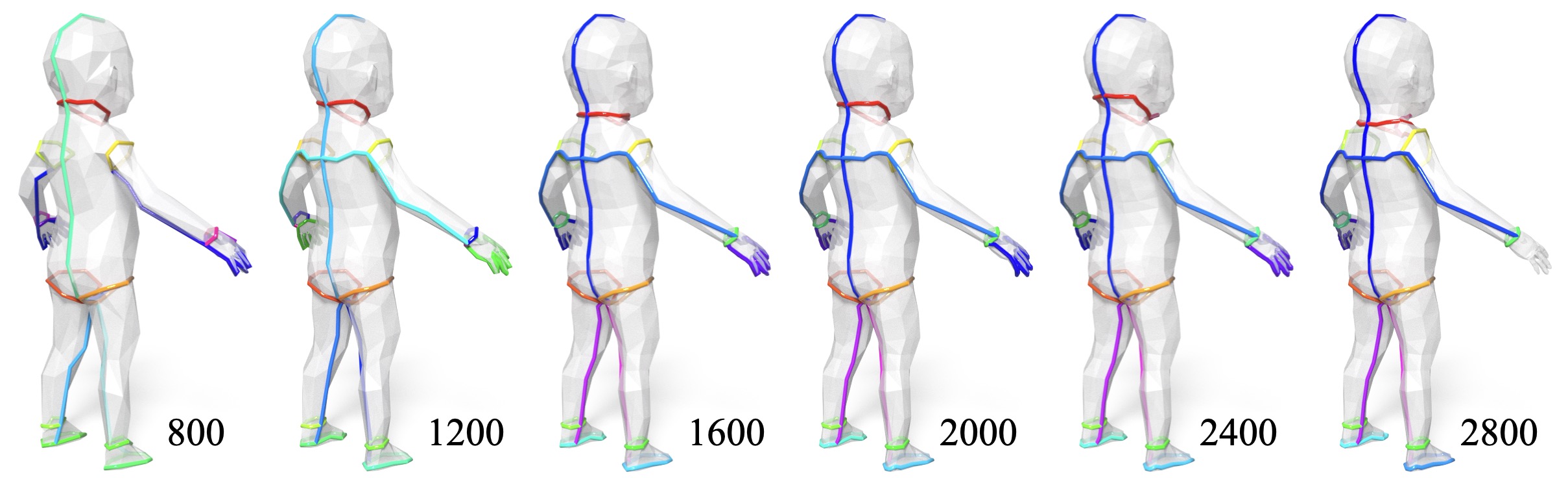}
  \caption{MeshTailor generalizes to meshes at different levels of granularity, handling meshes with more triangles than those seen during training.}
  \label{fig:multi_res}
\end{figure}

\paragraph{Generalization across mesh resolutions.}
The predicted seams retain a consistent global structure across mesh resolutions, even without re-training.
Although our model is trained on meshes with at most 2,000 triangles, Fig.~\ref{fig:multi_res} applies MeshTailor to meshes with increasing triangle counts (800 $\rightarrow$ 2,800); major loops and long-range cuts remain in similar locations with coherent connectivity.
Higher granularity mainly affects fine-scale details (e.g., slight boundary shifts) rather than the high-level cut layout, consistent with our coarse-to-fine ChainingSeams curriculum that places global cuts before local details and effectively decouples cut topology from mesh resolution.

\begin{figure}[t]
  \centering
  \includegraphics[width=\linewidth]{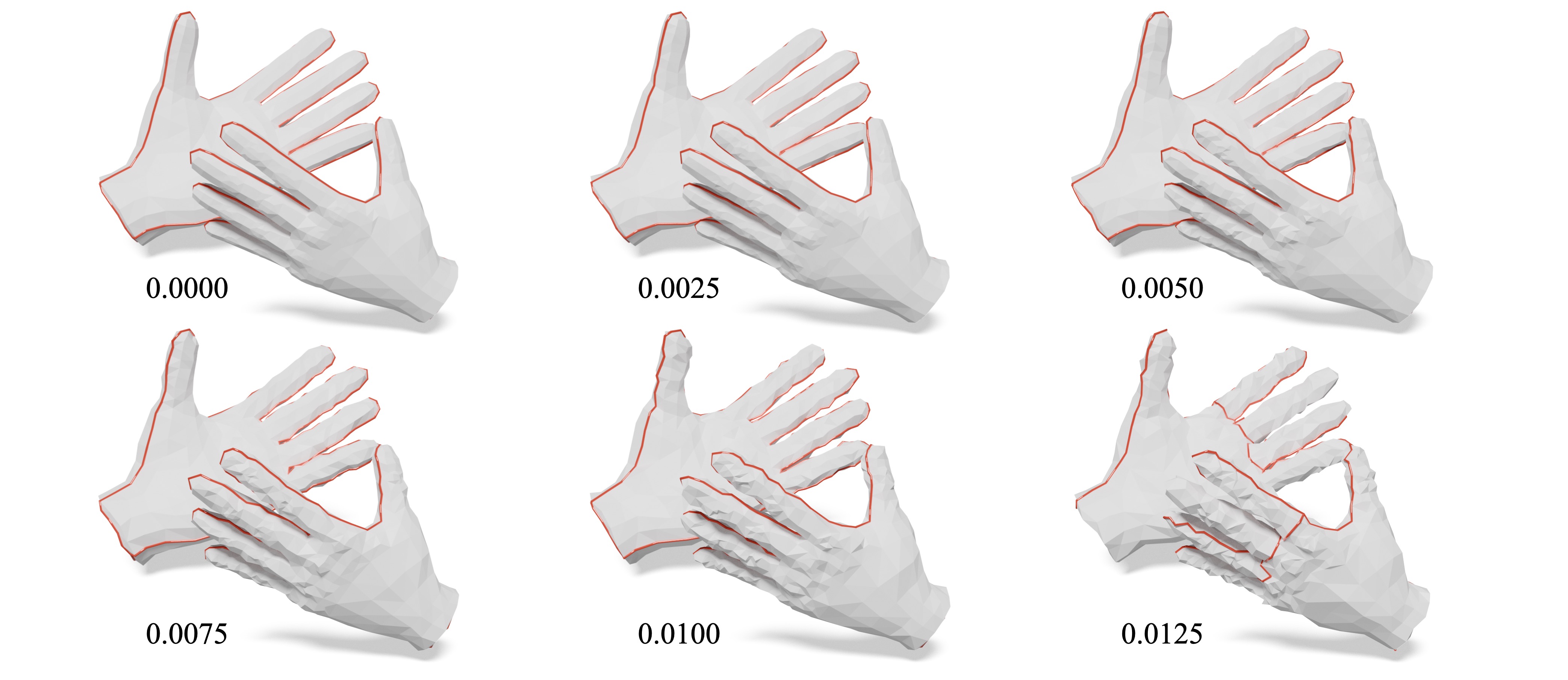}
  \caption{
  We perturb the input mesh by adding Gaussian noise to vertices with increasing $\sigma$.
  Predicted seams remain stable when $\sigma < 0.010$ and gradually degrade as $\sigma$ increases, with noticeable extra local cuts at $\sigma = 0.0125$.
  }
  \label{fig:noise_mesh}
\end{figure}

\paragraph{Stability under vertex jitter.}
Small vertex perturbations leave the predicted seams essentially unchanged, supporting reproducible seam decoding in practical asset pipelines.
Production assets frequently undergo minor geometry edits (e.g., smoothing or sculpting touch-ups) that perturb vertex positions while keeping connectivity unchanged; to probe stability in this regime, we apply isotropic Gaussian jitter to vertex positions.
Fig.~\ref{fig:noise_mesh} shows that MeshTailor produces nearly identical seams under perturbations with $\sigma \leq 0.01$, but may introduce a few extra cuts when $\sigma = 0.0125$ as geometric cues become less reliable.

\begin{figure*}
    \centering
    \includegraphics[width=0.95\linewidth]{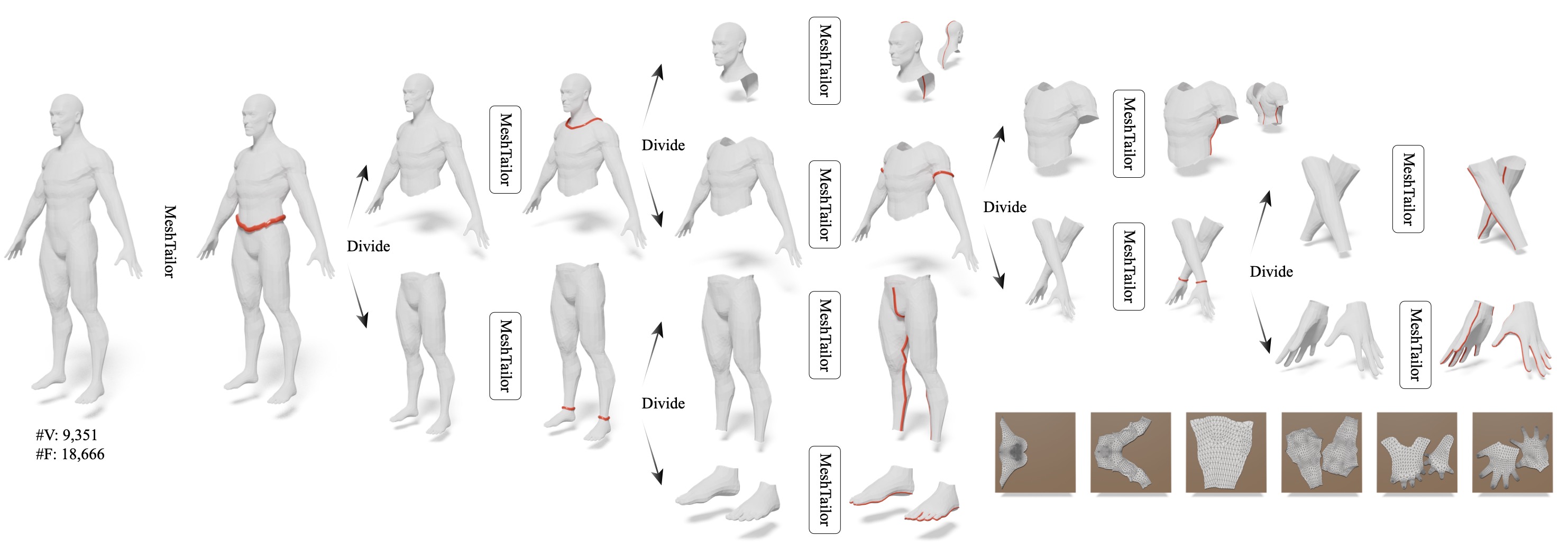}
    \caption{
    \textbf{Divide-and-conquer inference.}
    Starting from a high-resolution mesh, MeshTailor predicts a major seam loop to split the surface into disconnected components, then recursively applies the same seam-tracing process to each sub-mesh. This recursion progressively decomposes the asset into semantically coherent parts, yielding clean, edge-aligned seams and the corresponding UV charts.
    \vspace{-1mm}  
    }
    \label{fig:divide_and_conquer}
\end{figure*}

\paragraph{Optional divide-and-conquer decoding for dense meshes.}
For meshes whose triangle count substantially exceeds our training resolution, we optionally adopt a divide-and-conquer decoding scheme (Fig.~\ref{fig:divide_and_conquer}).
After a seam chain is generated, we test whether it splits the current patch into two connected components; if so, we recursively decode seams on each sub-mesh.
This optional mode preserves mesh-aligned seam continuity while shortening each decoding pass; we use it only when single-pass decoding on dense meshes becomes inconvenient, reusing the same model on each sub-mesh.

\subsection{Ablation Studies}
\label{sec:ablation}

\begin{figure}[t]
  \centering
  \includegraphics[width=\linewidth]{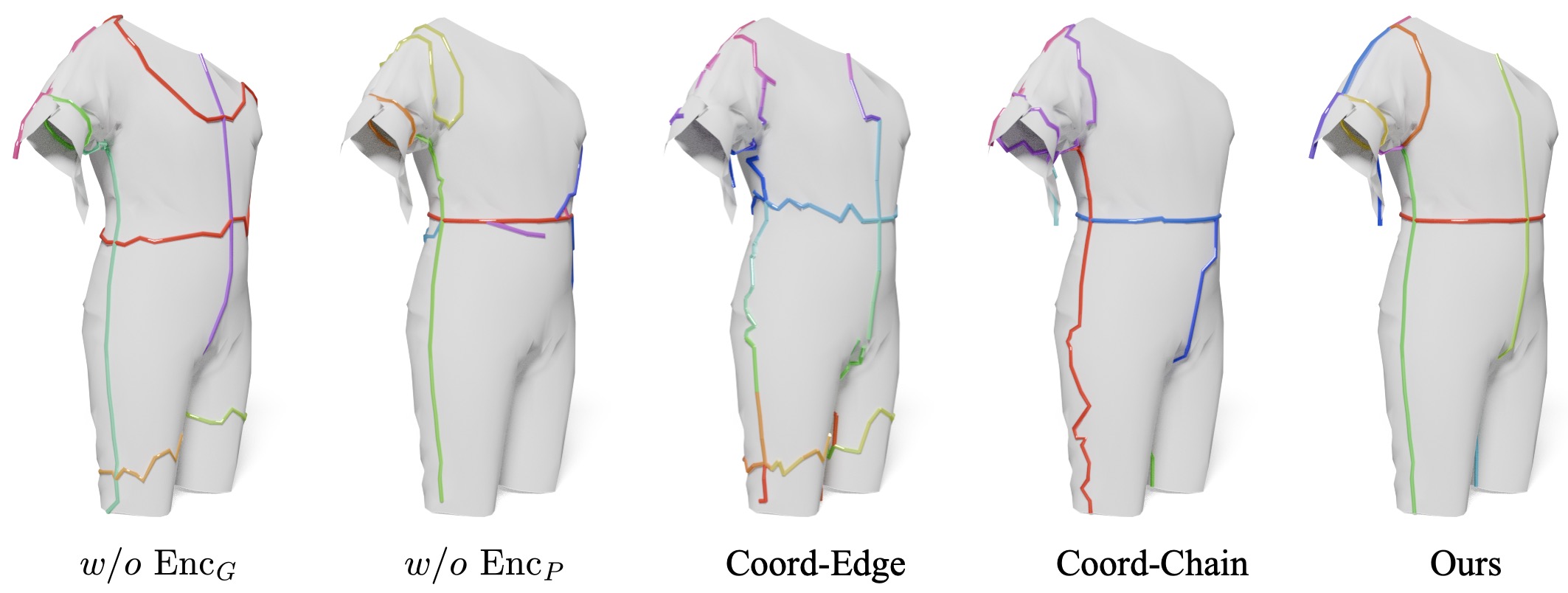}
  \caption{
    \textbf{Ablation on seam generation.} Encoder ablations (w/o $\mathrm{Enc}_G$, w/o $\mathrm{Enc}_P$) and coordinate-based baselines (Coord-Edge/Coord-Chain) versus our full method. Coordinate prediction yields irregular boundaries from required snapping.
  }
  \label{fig:ablation}

  \vspace{10pt}

  \captionof{table}{\textbf{Ablation study on GarmentCodeData.}
  We report complementary metrics covering distortion, chart complexity, and seam editability.
  Top-2 are highlighted as \colorbox{rankonecolor}{best} and \colorbox{ranktwocolor}{second}.
  }
  \label{tab:ablation}
  \scriptsize
  \setlength{\tabcolsep}{2.5pt}
  \renewcommand{\arraystretch}{1.0}

  \begin{tabular}{lccccccc}
  \toprule
  \shortstack[c]{\strut Variant\\[-2pt]\strut}
  & \shortstack{Overall\\Dist$\downarrow$}
  & \shortstack{Angular\\Dist$\uparrow$}
  & \shortstack{\#\\Charts$\downarrow$}
  & \shortstack{Island\\Compact$\uparrow$}
  & \shortstack{Island\\Convex$\uparrow$}
  & \shortstack{SeamLen\\/Area$\downarrow$}
  & \shortstack{Boundary\\Jagged$\downarrow$} \\
  \midrule
  w/o $\mathrm{Enc}_G$
  & \ranktwo1.125 & \rankone0.972 & 10.406 & 0.548 & \ranktwo0.865 & 2.292 & 0.539\\
  w/o $\mathrm{Enc}_P$
  & 1.129 & 0.969 & 11.281 & \ranktwo0.554 & 0.858 & 2.316 & \ranktwo0.538 \\
  \midrule
  Coord-Edge
  & 1.151 & 0.968 & \rankone9.287 & 0.442 & 0.844 & 2.325 & 1.074\\
  Coord-Chain
  & 1.135 & \ranktwo0.971 & \ranktwo9.987 & 0.496 & 0.842 & \ranktwo2.281 & 0.997\\
  \midrule
  Full model
  & \rankone\textbf{1.097} & 0.970 & 10.414 & \rankone\textbf{0.591} & \rankone\textbf{0.887} & \rankone\textbf{2.25} & \rankone\textbf{0.485}\\
  \bottomrule
  \end{tabular}

  \vspace{-2mm}
\end{figure}

Removing either encoder stream or replacing our mesh-native decoder with coordinate-based generation degrades the metrics in Table~\ref{tab:ablation}, confirming that all three components are necessary.
Without $\mathrm{Enc}_G$ (row 1), the decoder must infer connectivity from vertex positions alone; islands become less compact and convex, and boundaries more jagged.
Without $\mathrm{Enc}_P$ (row 2), the decoder loses the global shape prior, so the chart count grows and the layout becomes less regular (Fig.~\ref{fig:ablation}); shape semantics help merge fragments into fewer islands.
$\mathrm{Enc}_G$ thus captures local connectivity, $\mathrm{Enc}_P$ global semantics, jointly informing every pointer decision.

For the decoder, we reimplement coordinate-based generation from SeamGPT~\cite{SeamGPT} (no public code available) as two variants (rows 3-4): \emph{Coord-Edge} predicts seam-edge endpoints and \emph{Coord-Chain} predicts chain vertices, both quantizing each axis into 128 bins.
Unlike our pointer decoder, which is edge-aligned by construction, the coordinate variants must snap predictions to the nearest mesh vertex, producing substantially more jagged boundaries (Fig.~\ref{fig:ablation}).
The formulation also lengthens the output sequence by $3.4\times$ to $5.1\times$ on our test set and expands the per-step output space from 5-6 neighboring vertices to 128 bins, making learning harder.

\section{Conclusion and Future Work}
\label{sec:conclusion}

We have presented MeshTailor, the first mesh-native generative framework for synthesizing edge-aligned UV seams directly on 3D mesh graphs. 
By operating intrinsically on mesh topology, our approach eliminates projection artifacts and guarantees edge-aligned, continuous seam boundaries by construction. 
Our key innovations are three: ChainingSeams, a serialization scheme that decomposes the unordered seam graph into ordered vertex walks; a dual-stream encoder that fuses local mesh connectivity with global shape semantics; and a mesh-native pointer layer that constrains autoregressive decoding to mesh edges.
Extensive evaluations show that MeshTailor produces seam layouts close to artist-authored layouts, with more regular, edge-aligned islands than greedy, optimization-based, and recent neural baselines.

\paragraph{Limitations and Future Work.}
Our framework currently focuses on low-polygon meshes; higher-resolution inputs can be addressed by scaling up the model or, optionally, by applying our divide-and-conquer inference.
Future directions include interactive seam editing with user constraints, distortion-aware training objectives, and applying mesh-native graph traversal to other decomposition tasks such as remeshing and mesh segmentation.

\clearpage

{
    \small
    \bibliographystyle{ieeenat_fullname}
    \bibliography{SeamGen}
}


\clearpage
\appendix

\twocolumn[{%
\begin{center}
{\LARGE\textbf{MeshTailor: Cutting Seams via Generative Mesh Traversal}}\\[1em]
{\Large Supplemental Material}\\[0.5em]

{\large
Xueqi Ma\textsuperscript{1} \quad
Xingguang Yan\textsuperscript{2} \quad
Congyue Zhang\textsuperscript{1} \quad
Hui Huang\textsuperscript{1*}
}\\[0.2em]
{\large
\textsuperscript{1}Shenzhen University \quad
\textsuperscript{2}Simon Fraser University
}
\end{center}
}]

\let\thefootnote\relax\footnotetext{
$^*$Corresponding author.
}

This supplementary material complements the main paper with additional technical and experimental details.
It includes: 
(i) dataset preprocessing and seam extraction from UV layouts (Sec.~\ref{sec:supp_data}); 
(ii) extended method details on seam-chain representation, canonical ordering, and architecture (Sec.~\ref{sec:supp_representation}--\ref{sec:supp_arch}); 
(iii) complete evaluation protocols, including metric definitions, user study (Sec.~\ref{sec:supp_metrics}--\ref{sec:supp_user_study}); 
and (iv) additional visualizations of distortion and UV layouts (Sec.~\ref{sec:supp_uv}).
We also provide further analysis of coordinate-based decoding baselines (Sec.~\ref{sec:supp_coord}), additional qualitative results (Sec.~\ref{sec:additional_results}), and representative failure cases and limitations (Sec.~\ref{sec:supp_failure}).

\section{Data}
\subsection{Data Processing}
\label{sec:supp_data}

\paragraph{Dataset sources.}
We train and evaluate on two large-scale datasets covering diverse assets:
TexVerse~\cite{TexVerse} and GarmentCodeData~\cite{GarmentCodeData}.

\begin{figure}[t]
    \centering
    \includegraphics[width=0.95\linewidth]{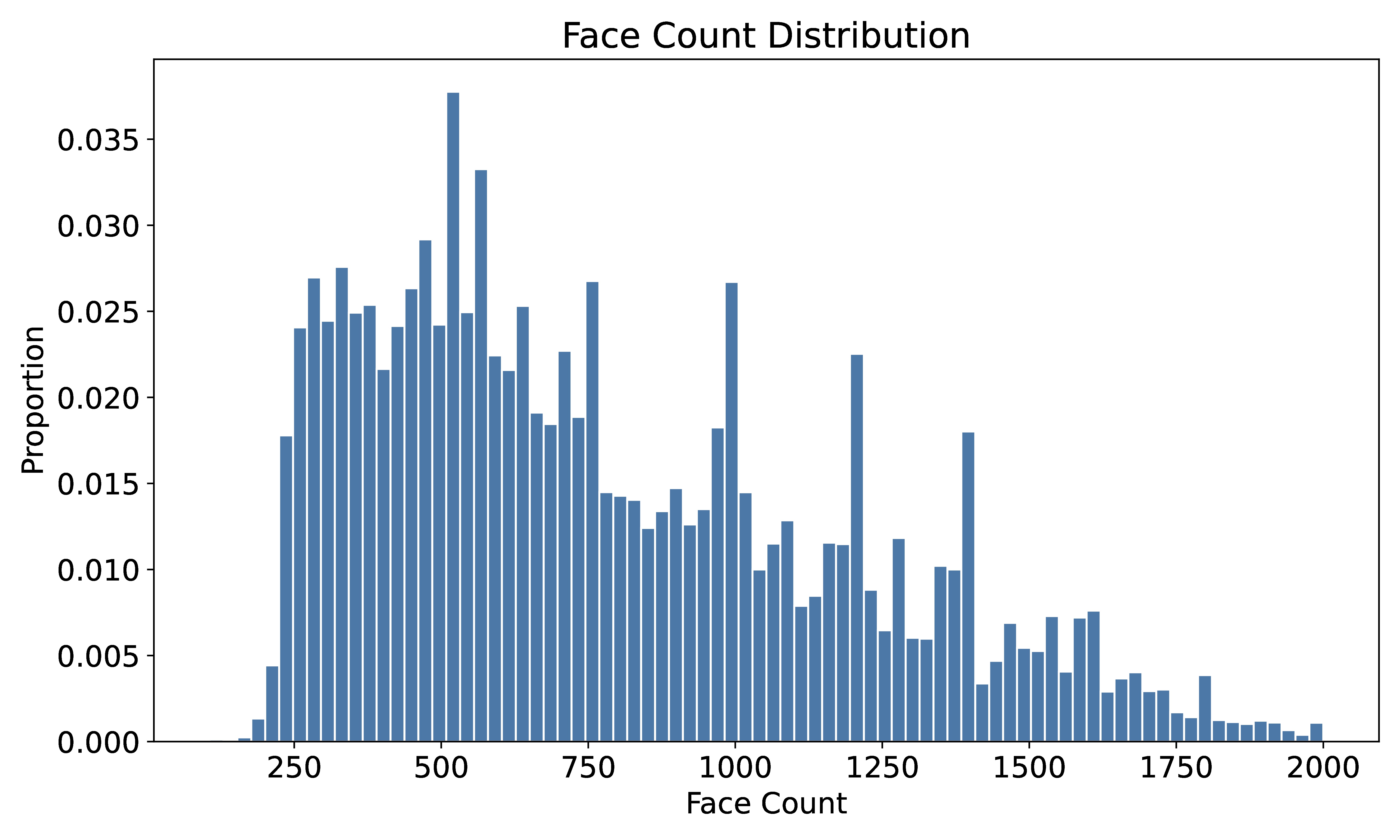}
    \caption{
    \textbf{Triangle-count distribution of TexVerse part-level training samples.}
    Histogram of triangle counts across the part-level samples retained after the preprocessing rules described in Sec.~\ref{sec:supp_data}.
    }
    \label{fig:face_count_distribution}
\end{figure}

\paragraph{TexVerse preprocessing.}
For TexVerse, we first filter for high-quality UV layouts by checking for valid per-corner UV coordinates and ensuring all faces have consistent UV mappings.
We then decompose each mesh into part-level connected components to match production assets where seams are authored per part.
We cap the training resolution at 2,000 triangles (Fig.~\ref{fig:face_count_distribution}), since the majority of part-level samples already fall below this threshold; meshes exceeding the limit are decimated \cite{Seamless} to below 2,000 while preserving geometric detail.
Finally, we convert meshes to manifold surfaces using standard mesh repair operations to ensure consistent topology.

\paragraph{VLM-based filtering.}
We further remove overly simple primitives (e.g., cuboids, cylinders, spheres) using a VLM (Qwen3-VL~\cite{Qwen3-VL}).
These cases provide little seam-structure diversity and can dominate training, biasing the model toward trivial solutions.
The VLM is prompted to classify whether a mesh is a simple geometric primitive based on a $2\times2$ grid of four rendered views.

\paragraph{Seam extraction from UV layouts.}
Given per-corner UV coordinates, we extract seam edges as mesh edges whose shared endpoints are \emph{not UV-glued} across the two incident faces (up to orientation).
Specifically, for each mesh edge $e=(v_a, v_b)$ shared by two faces $f_1$ and $f_2$, we check whether the UV coordinates of $v_a$ and $v_b$ differ across $f_1$ and $f_2$.
If they differ, the edge is marked as a seam edge.
In practice, we use Blender's seam extraction operator \texttt{bpy.ops.uv.seams\_from\_islands()}~\cite{blender_seams_from_islands} to obtain seam edges $\mathcal{S}$.

\section{Method Details}

\subsection{Detailed Seam Chain Representation}
\label{sec:supp_representation}

\paragraph{Seam as an edge set.}
In UV unwrapping, seams indicate mesh edges along which the surface is cut to produce one or multiple UV charts.
We formalize a seam layout as an edge subset $\mathcal{S}\subseteq\mathcal{E}$.
Equivalently, each edge $e\in\mathcal{E}$ can be assigned a binary label
$y_e\in\{0,1\}$ with $y_e=\mathbb{I}[e\in\mathcal{S}]$. 
However, treating seams as independent edge labels under-specifies their \emph{global structure}.
In practice, seams must exhibit coherent connectivity (continuity, frequent loop structures, and strong coupling to the underlying triangulation).
To model such structure explicitly, we adopt a serialized \emph{seam chain} representation compatible with autoregressive generation.

\paragraph{Seam chain.}
We represent a seam layout as a set of seam chains $\mathcal{C}$.
Training targets are constructed from the extracted seam-edge set $\mathcal{S}$ (Sec.~\ref{sec:supp_data}). We decompose $\mathcal{S}$ into chains $\mathcal{C}$ by tracing maximal edge-connected paths on the seam subgraph, yielding both open chains and closed loop cuts.
Formally, a seam chain is a vertex sequence
$\mathbf{c}=(v_1,v_2,\ldots,v_T)$ with $v_t\in\mathcal{V}$ that forms a walk on the mesh:
\begin{equation}
(v_t,v_{t+1})\in\mathcal{E},\quad \forall t\in\{1,\ldots,T-1\}.
\label{eq:chainwalk}
\end{equation}
It induces an ordered seam-edge list $\{(v_t,v_{t+1})\}_{t=1}^{T-1}$.
We denote a \emph{loop cut} as a special chain whose start and end coincide, i.e., $v_1=v_T$.
Loop cuts are prevalent in artist seam annotations (e.g., rings around limbs or torso junctions),
and our chain formalism supports them naturally in practice.

\paragraph{Mesh-native validity via neighbor masking.}
Eq.~\eqref{eq:chainwalk} implies a key constraint: the next vertex must be adjacent to the current vertex.
In our decoder, we enforce this \emph{mesh-native} constraint by masking the pointer distribution to the 1-ring neighbors of the current vertex,
ensuring that every generated chain is edge-aligned by construction.

\subsection{Detailed Formulation of Canonical Ordering}
\label{sec:supp_ordering}

\paragraph{Formalization.}
We formalize the canonical ordering introduced in Sec.~\ref{sec:chainingseams} as an iterative divide process: at each step, we split the current largest patch by the loop cut that yields the most balanced sub-patches, then recurse on the resulting sub-patches.

\textbf{Problem setting.}
Let $\mathcal{P}$ be a set of disjoint connected patches, initialized as $\mathcal{P}\leftarrow\{\mathcal{M}\}$.
Let $\mathcal{L}$ be the set of loop cuts in $\mathcal{C}$, and $\mathcal{O}=\mathcal{C}\setminus\mathcal{L}$ the set of open chains.
At each iteration, we select a patch to refine and choose the next loop within that patch.

\textbf{Loop candidates within a patch.}
For a patch $P\in\mathcal{P}$, we collect internal loop candidates
\begin{equation}
\mathcal{L}_{P}=\{L\in\mathcal{L}\mid L\subset P,\; L \text{ is not coincident with }\partial P\}.
\end{equation}
Cutting $P$ along $L\in\mathcal{L}_{P}$ splits $P$ into two connected sub-patches
$P^{(1)}_L$ and $P^{(2)}_L$ with areas $A^{(1)}_L=\mathrm{Area}(P^{(1)}_L)$ and
$A^{(2)}_L=\mathrm{Area}(P^{(2)}_L)$.

\textbf{Area-balance score.}
We define the balance score
\begin{equation}
r(L;P)=\frac{\min(A^{(1)}_L,A^{(2)}_L)}{\max(A^{(1)}_L,A^{(2)}_L)}\in(0,1],
\label{eq:balance}
\end{equation}
where values closer to $1$ indicate a more even split.
We choose the loop with maximal balance within the selected patch.

\textbf{Patch selection.}
We always refine the patch with the largest area at each iteration:
\begin{equation}
P^\star=\arg\max_{P\in\mathcal{P}} \mathrm{Area}(P).
\label{eq:patchpick}
\end{equation}
Intuitively, large patches dominate potential distortion and visual impact, motivating our largest-patch-first selection.

\textbf{Iterative ordering.}
We repeatedly (i) select $P^\star$ by Eq.~\eqref{eq:patchpick}, 
(ii) pick the next loop cut $L^\star$ as
\begin{equation}
L^\star=\arg\max_{L\in \mathcal{L}_{P^\star}} r(L;P^\star),
\end{equation}
(iii) append $L^\star$ to the ordered chain list, and (iv) split $P^\star$ into the two sub-patches induced by $L^\star$ and update $\mathcal{P}$.
If $\mathcal{L}_{P^\star}=\emptyset$, we remove $P^\star$ from $\mathcal{P}$ and continue with the next-largest patch.
We remove $L^\star$ from $\mathcal{L}$ after each selection, ensuring each loop cut is used at most once.
After all loop cuts are ordered, we append the open chains $\mathcal{O}$ in decreasing order of chain length, yielding a complete deterministic sequence.
This ordering converts the original set $\mathcal{C}$ into an autoregressive supervision signal and makes target construction deterministic (Algorithm~\ref{alg:chain_order}).

\begin{algorithm}[t]
\caption{Canonical ordering of seam chains: loops first, largest patch first, area balance.}
\label{alg:chain_order}
\SetAlgoNoLine \DontPrintSemicolon
\KwIn{mesh $\mathcal{M}$; seam chains $\mathcal{C}$}
\KwOut{ordered chain list $\mathrm{Seq}$}

$\mathcal{L}\leftarrow\{\,\mathbf{c}\in\mathcal{C}\mid \mathbf{c}\ \text{is a loop cut}\,\}$\;
$\mathcal{O}\leftarrow\mathcal{C}\setminus\mathcal{L}$\tcp*{open chains}

$\mathcal{P}\leftarrow\{\mathcal{M}\}$;\quad $\mathrm{Seq}\leftarrow[\,]$\tcp*{patches / ordered}

\While{$\mathcal{P}\neq\emptyset$}{
  $P^\star \leftarrow \arg\max_{P\in\mathcal{P}}\mathrm{Area}(P)$\tcp*{largest patch}
  $\mathcal{L}_{P^\star}\leftarrow\{\,L\in\mathcal{L}\mid L\subset P^\star,\ L\not\subset\partial P^\star\,\}$\tcp*{internal loops}

  \If{$\mathcal{L}_{P^\star}=\emptyset$}{
    $\mathcal{P}\leftarrow\mathcal{P}\setminus\{P^\star\}$;\;
    \textbf{continue}\;
  }

  $L^\star \leftarrow \arg\max_{L\in\mathcal{L}_{P^\star}} r(L;P^\star)$\tcp*{Eq.~\eqref{eq:balance}}
  $\mathrm{Seq}.\mathrm{append}(L^\star)$\;

  $(P_1,P_2)\leftarrow \mathrm{Split}(P^\star,L^\star)$\tcp*{cut $P^\star$ along $L^\star$}
  $\mathcal{P}\leftarrow(\mathcal{P}\setminus\{P^\star\})\cup\{P_1,P_2\}$\;
  $\mathcal{L}\leftarrow\mathcal{L}\setminus\{L^\star\}$\tcp*{use each loop once}
}

$\mathrm{Seq}.\mathrm{extend}\big(\mathrm{Sort}(\mathcal{O})\big)$\tcp*{open chains: by decreasing length}

\Return $\mathrm{Seq}$\;
\end{algorithm}

\subsection{Architecture}
\label{sec:supp_arch}
This section details the architecture of our method, a mesh-native seam generator that predicts seam chains as an autoregressive stream of \emph{vertex selections}, with two special symbols \texttt{[EOC]} and \texttt{[EOS]} indicating end-of-chain and end-of-sequence, respectively.

\subsubsection{Inputs, Tokenization, and Candidate Space} 

Given a mesh $\mathcal{M}=(\mathcal{V},\mathcal{E},\mathcal{F})$ with $|\mathcal{V}|=N$ vertices, the network additionally takes a point cloud
$\mathbf{P}\in\mathbb{R}^{N_p\times 6}$ sampled on the surface, where each point stores $(x,y,z)$ and its normal $(n_x,n_y,n_z)$.
We represent the target seam layout as an index sequence
$\{\tau_t\}_{t=1}^{T}$, where each $\tau_t$ is either a vertex index in $\{0,\dots,N-1\}$ or one of the special symbols
\texttt{[EOC]} / \texttt{[EOS]}.

To unify pointer decoding, we construct a \emph{candidate set} that contains all vertices plus two special candidates.
In our implementation, we place the two special candidates before vertices:
\[
\mathcal{U} = \{\texttt{[EOC]}, \texttt{[EOS]}\} \cup \mathcal{V},
\]
and use a consistent offset mapping so that vertex indices are shifted by $+2$ in the candidate space.
Concretely, we map a vertex id $v\in[0,N\!-\!1]$ to candidate id $(v+2)$, while \texttt{[EOC]} and \texttt{[EOS]} are mapped to candidate ids $0$ and $1$, respectively.

\subsubsection{Encoding}

We compute the final per-vertex embeddings in five steps:
(i) coordinate-based vertex features,
(ii) mesh connectivity encoding,
(iii) raw-coordinate fusion,
(iv) global shape token extraction,
and (v) shape-to-vertex cross-attention.

\paragraph{Coordinate-based vertex features.}
Each per-vertex feature $\mathbf{v}_i\in\mathbb{R}^6$ (position concatenated with normal) is first embedded by a Fourier feature encoder $\phi(\cdot)$.
We concatenate the embedding with the raw feature and project it into a learnable per-vertex feature:
\[
\mathbf{p}_i=\mathrm{MLP}\big([\phi(\mathbf{v}_i),\ \mathbf{v}_i]\big)\in\mathbb{R}^{d_p},
\]
where $d_p=384$ in our default setting.

\paragraph{Mesh connectivity encoder.}

\begin{figure}[t]
  \centering
  \includegraphics[width=\linewidth]{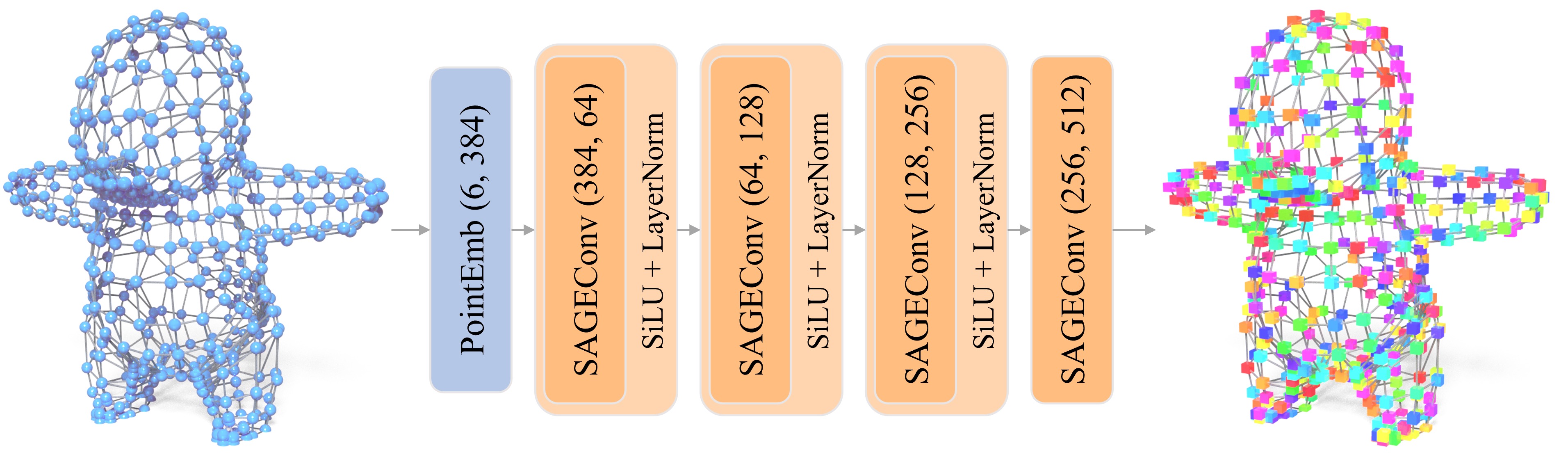}
  \caption{
    \textbf{Graph Encoder.}
    We embed each vertex with coordinate-based point features and apply a stack of GraphSAGE layers with SiLU and LayerNorm to aggregate neighborhood information, producing per-vertex connectivity-aware embeddings used for seam generation.
  }
  \label{fig:graph_encoder}
\end{figure}

To explicitly capture mesh-dependent seam decisions, we encode mesh connectivity with a multi-layer GraphSAGE encoder (Fig.~\ref{fig:graph_encoder}).
We perform message passing over the mesh edges $\mathcal{E}$.
Starting from the coordinate-based vertex features $\mathbf{p}_i$, we stack multiple GraphSAGE layers:
\[
\mathbf{h}_i^{(g,\ 0)}=\mathbf{p}_i,\qquad
\mathbf{h}_i^{(g,\ \ell+1)}=\mathrm{SAGEConv}^{(\ell)}\big(\mathbf{h}^{(g,\ \ell)},\mathcal{E}\big)_i,
\]
where each layer is followed by $\mathrm{SiLU}$ and $\mathrm{LayerNorm}$ for stable optimization.
In our default setting, the hidden widths are set to \texttt{[64, 128, 256, 512]}, producing the final connectivity-aware vertex embedding $\mathbf{h}^g_i\in\mathbb{R}^{512}$.

\paragraph{Raw-coordinate fusion.}
We concatenate the final graph feature with the coordinate-based vertex feature $\mathbf{p}_i$ and project them to the model dimension $d$:
\[
\mathbf{h}_i = \mathbf{W}_f\,[\mathbf{h}_i^{g},\ \mathbf{p}_i]\in\mathbb{R}^{d},
\]
with $d=512$.
This fusion is intended to retain accurate coordinate cues alongside connectivity-aware features.

\paragraph{Pretrained shape encoder.}
We extract global shape semantics from the surface point cloud using a pretrained point-cloud encoder~\cite{Michelangelo}.
It outputs a set of shape tokens $\mathbf{Z}\in\mathbb{R}^{M\times d}$, where $d$ matches the model dimension.
During training, the shape encoder parameters are frozen by default.

\paragraph{Shape-to-vertex cross-attention.}

Finally, our cross-attention fusion takes vertex embeddings as queries and shape tokens as key/value, injecting global shape context into the local vertex representations:
\[
\mathbf{\tilde{h}}_i=\mathrm{CrossAttn}\ \big(\mathbf{h}_i,\ \mathbf{Z}\big),
\]
where we use a small depth (2 layers by default) to keep the overhead minimal while enabling global context aggregation.
The resulting $\mathbf{\tilde{h}}_i\in\mathbb{R}^{d}$ serves as the final per-vertex token representation.

\subsubsection{Autoregressive Decoder}

Given vertex tokens $\{\mathbf{\tilde{h}}_i\}$, our decoder predicts the next element in the seam stream by pointing to candidates in $\mathcal{U}$.

\paragraph{Candidate Embeddings.}

We build candidate embeddings by concatenating two learned special embeddings with the per-vertex embeddings:
\[
\mathbf{Cands} = [\mathbf{e}_{\texttt{[EOC]}},\ \mathbf{e}_{\texttt{[EOS]}},\ \mathbf{\tilde{h}}_0,\dots, \mathbf{\tilde{h}}_{N-1}]\in\mathbb{R}^{(N+2)\times d}.
\]
A candidate mask is built accordingly, marking valid vertices and always allowing the special candidates.

\paragraph{Decoder input and positional encoding.}
We use a decoder-only Transformer with 6 layers and model dimension 512, and apply rotary positional embeddings (RoPE) ~\cite{rope} to encode token order.
In addition, we introduce a chain-local positional embedding that resets after every \texttt{[EOC]} to encode within-chain progress.
Let $\pi_t$ denote the number of steps since the most recent \texttt{[EOC]} (starting from $0$ at each chain start).
We add
\[
\tilde{\mathbf{e}}_t \leftarrow \mathbf{e}_t + \mathbf{PosEmb}^{\text{chain}}_{\pi_t},
\]
which helps the model distinguish early versus late vertices inside the current chain, independently of the global decoding index.

\paragraph{Conditioned decoding on shape context.}
At each step, the decoder attends to the previously generated token embeddings and is conditioned on the shape context tokens $\mathbf{Z}$.
Denoting the decoder hidden state at step $t$ as $\mathbf{q}_t\in\mathbb{R}^{d}$, we compute it as
\[
\mathbf{q}_t=\mathrm{Decoder}(\tilde{\mathbf{e}}_{\le t};\ \mathbf{Z}),
\]
which improves long-range decisions such as globally meaningful loop placement and cross-part consistency (cf.\ \emph{w/o $\mathrm{Enc}_P$} in Sec.~\ref{sec:ablation}).

\paragraph{Pointer projection.}
Let $\mathbf{e}_u := \mathbf{Cands}_u$ denote the embedding of candidate $u$ (the $u$-th row of $\mathbf{Cands}$). We compute logits over candidates via dot-product attention:
\[
\ell_{t,u}=\langle \mathbf{q}_t, \mathbf{W}\,\mathbf{e}_u \rangle.
\]

\subsubsection{Neighbor-Constrained Decoding at Inference}

During inference, we restrict the pointer choices using a dynamic candidate mask to enforce edge-aligned traversal and prevent degenerate jumps.
We perform autoregressive sampling with temperature $0.1$, fixed across all experiments.
We precompute a 1-ring neighbor table for each vertex and apply the following rules based on the last predicted token $\tau_t$:

\begin{itemize}
\item If $\tau_t=\texttt{[EOC]}$, we start a new chain and only allow valid vertices as the next token.
\item If $\tau_t$ is a vertex, we only allow its 1-ring neighbors as candidates and additionally allow \texttt{[EOC]} and \texttt{[EOS]}.
To avoid immediate backtracking, we remove the previous vertex from the neighbor set when applicable.
\item If $\tau_t=\texttt{[EOS]}$, generation terminates.
\end{itemize}

This neighbor-constrained mask makes every decoded step mesh-native by construction.
It also reduces the per-step candidate set from $O(N)$ to a small local subset, which keeps decoding efficient.

\subsubsection{Inference Pseudocode}
\label{sec:supp_infer_pseudocode}
Algorithm~\ref{alg:inference} summarizes the inference-time decoding procedure of MeshTailor.
Starting from the encoded mesh and shape features, the model generates seam chains in an autoregressive manner by repeatedly selecting vertices under a dynamic neighbor mask.
This masking enforces edge-aligned traversal, ensuring mesh-native generation without post-hoc projection.

\begin{algorithm}[t]
\caption{MeshTailor inference}
\label{alg:inference}
\SetAlgoNoLine
\KwIn{Manifold mesh $\mathcal{M}=(\mathcal{V},\mathcal{E},\mathcal{F})$}
\KwOut{Seam chains $\mathcal{C}$ (edge-aligned by construction)}
\BlankLine

$\{\mathbf{h}_i\}\leftarrow \mathrm{Enc}_G(\mathcal{M})$; $\mathbf{Z}\leftarrow \mathrm{Enc}_P(\mathbf{P})$\;
$\tilde{\mathbf{h}}_i \leftarrow \mathrm{CrossAttn}(\mathbf{h}_i,\mathbf{Z})$\;
Initialize empty chain list $\mathcal{C}$\;

\While{not \texttt{\upshape [EOS]}}{
    Compute start-vertex distribution over $\mathcal{V}$ (no neighbor mask)\;
    Select $v_1 \in \mathcal{V}$ by sampling\;
    Initialize current chain $\mathbf{c}\leftarrow (v_1)$\;

    \While{not \texttt{\upshape [EOC]}}{
        Compute pointer distribution with neighbor mask (Eq.~\ref{eq:pointer})\;
        Select next token $\tau$ by sampling\;
        Let $v_{t+1}\leftarrow \tau$ and append it: $\mathbf{c}\leftarrow (v_1,\ldots,v_{t+1})$\;
        $t \leftarrow t+1$\;
    }
    Append $\mathbf{c}$ to $\mathcal{C}$\;
}
\Return $\mathcal{C}$\;
\end{algorithm}

At inference, MeshTailor takes on average about 1.9 seconds per mesh at the training resolution ($\le$2k triangles) on a single NVIDIA RTX 4090.

\section{Experiments}
\subsection{Metric Details}
\label{sec:supp_metrics}

UV seams influence downstream parameterization along multiple (often competing) axes, e.g., distortion versus chart fragmentation and boundary regularity. Therefore, distortion alone is insufficient to fully reflect production usability. We evaluate along four complementary axes:

\paragraph{Distortion metrics.}
Following PartUV~\cite{partuv}, we report multiple distortion measures that quantify how UV mapping deforms the surface:
\begin{itemize}
\item \textbf{Overall distortion}: 
Measures global \emph{area} distortion by aggregating triangle-wise area stretch over the entire UV atlas.
\item \textbf{Angle preservation}: Measures the degree to which angles are preserved by the UV mapping; values closer to $1.0$ indicate closer-to-conformal mapping (higher is better).
\end{itemize}
These metrics reflect both area and angle preservation.

\paragraph{UV island quality metrics.}
To capture whether islands are \emph{regular and easy to pack/edit}, we measure:
\begin{itemize}
\item \textbf{Compactness}: Defined as $4\pi \cdot \mathrm{area}/\mathrm{perimeter}^2$. This metric ranges from 0 to 1, with 1 indicating a perfect circle. Higher values indicate more compact shapes.
\item \textbf{Convexity}: Island area divided by its convex-hull area. This ranges from 0 to 1,  where 1 is perfectly convex.
\end{itemize}
These metrics favor compact and convex shapes, penalizing elongated, highly concave, and jagged islands that complicate packing and manual editing in UV authoring tools.

\paragraph{Seam properties.}
We measure seam characteristics:
\begin{itemize}
\item \textbf{Seam length / 3D area}: 
Total seam length normalized by surface area. Lower values indicate more efficient cuts.
\item \textbf{Jaggedness/smoothness}: 
While seam regularity is defined on the cut graph in 3D, after unwrapping, the cut graph manifests as the \emph{UV chart boundaries} (i.e., island outlines). 
We therefore evaluate seam smoothness by measuring the boundary regularity of UV islands.
Concretely, we extract island boundary loops in the UV domain, resample each loop at uniform arc-length steps, and compute a discrete curvature proxy
$\kappa_i=\|\mathbf{p}_{i-1}-2\mathbf{p}_i+\mathbf{p}_{i+1}\|_2$ (cyclic indexing) on the resampled sequence $\{\mathbf{p}_i\}$.
We report the mean $\kappa$ over all boundary samples; lower values indicate smoother, less jagged seams.
\end{itemize}
These metrics directly reflect seam \emph{editability} and overall boundary regularity. 
Smoother seams are generally easier for artists to manipulate and adjust.

\paragraph{Structural complexity.}
We report the number of charts (UV islands) as a proxy for cut complexity. Fewer charts are generally preferable in production, provided that distortion and boundary quality remain acceptable. Fewer charts mean fewer discontinuities and easier texture painting workflows.

\begin{figure}[t]
    \centering
    \includegraphics[width=0.99\linewidth]{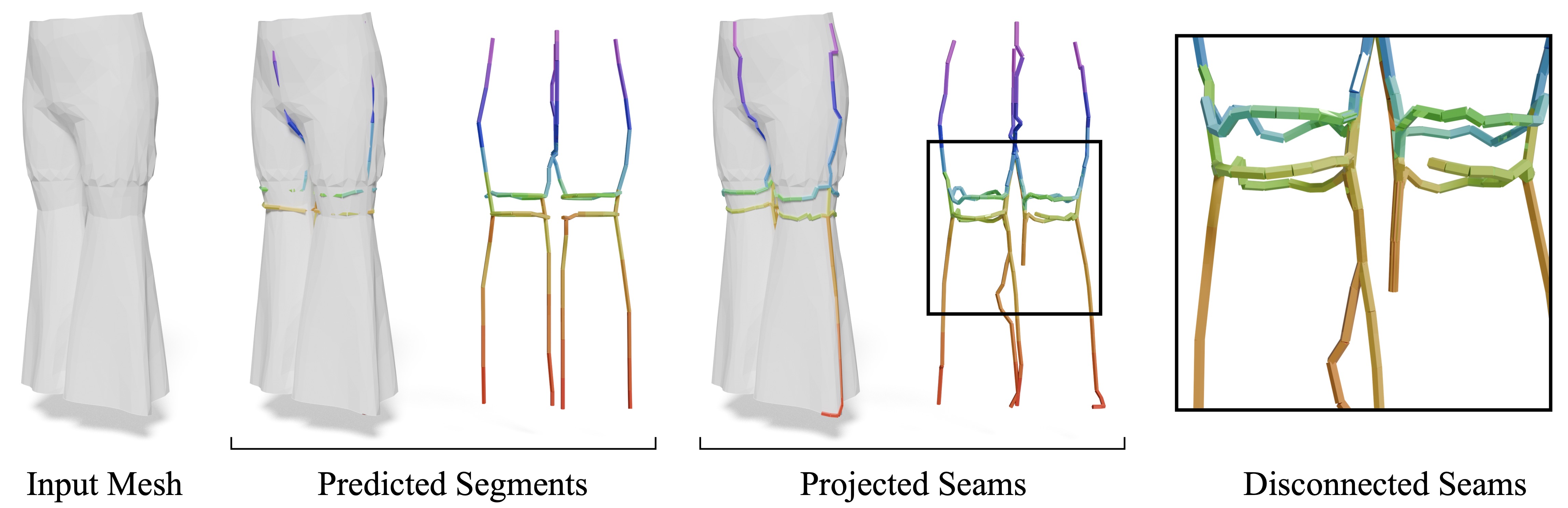}
    \caption{
    \textbf{Coordinate-to-mesh projection.}
    Coordinate-based seams can look plausible in Euclidean space, but snapping them to the input mesh often induces jagged boundaries and may collapse onto the wrong surface sheet under nearby/interpenetrating geometry, leading to disconnections.
    }
    \label{fig:coord_proj}
\end{figure}

\begin{figure}[t]
  \centering
  \includegraphics[width=\linewidth]{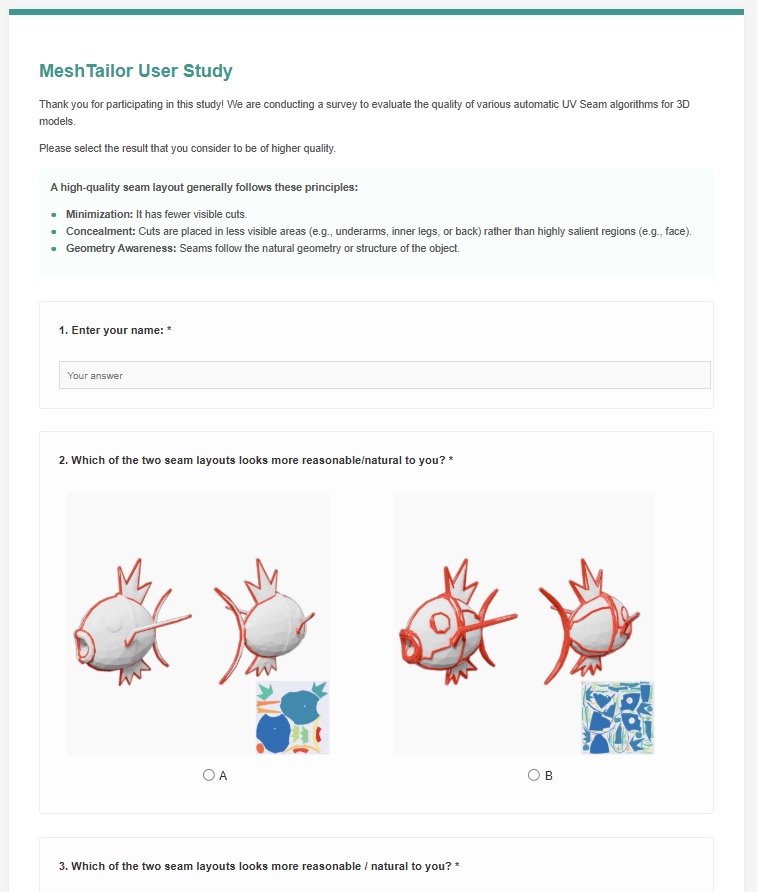}
    \caption{\textbf{User study interface.}
    We present participants with paired seam layout results (including seam visualizations on the 3D mesh and the corresponding UV maps) and ask them to choose the higher-quality option in a 2AFC setting, following the provided criteria (minimization, concealment, and geometry awareness).
    }
  \label{fig:user_study_interface}
\end{figure}

\subsection{User Study Details}
\label{sec:supp_user_study}

\paragraph{Study design.}
We conducted a two-alternative forced-choice (2AFC) user study comparing seam layouts across different methods.
Seam quality is partly governed by production conventions (e.g., visibility and editability) that are not fully captured by distortion metrics.

\paragraph{Participant recruitment.}
We recruited 100 participants through an online crowdsourcing platform. Participants were screened to ensure basic familiarity with 3D graphics and UV mapping concepts.

\paragraph{Study protocol.}
In each trial, participants were shown two results of the same 3D model processed by two different methods, including both the seam visualization (rendered on the 3D mesh) and the resulting UV map layout (Fig.~\ref{fig:user_study_interface}).

Participants selected the seam layout they considered higher quality, following three explicit criteria:
\begin{itemize}
\item \textbf{Minimization}: Fewer visually apparent cuts on the 3D surface.
\item \textbf{Concealment}: Cuts placed in less salient regions (e.g., underside, back, natural boundaries).
\item \textbf{Geometry awareness}: Seams following natural geometric or structural boundaries rather than arbitrary paths.
\end{itemize}

\paragraph{Study scale.}
Each participant completed 50 randomized pairwise comparisons.
Comparisons were balanced across method pairs to ensure each pair received approximately equal coverage.
In total, we collected responses from 100 participants, yielding 5,000 votes.

\subsection{UV Layout}
\label{sec:supp_uv}
\begin{figure*}[p]
    \centering
    \begin{minipage}[c][\textheight][c]{\textwidth}
        \centering
        \includegraphics[width=0.93\linewidth]{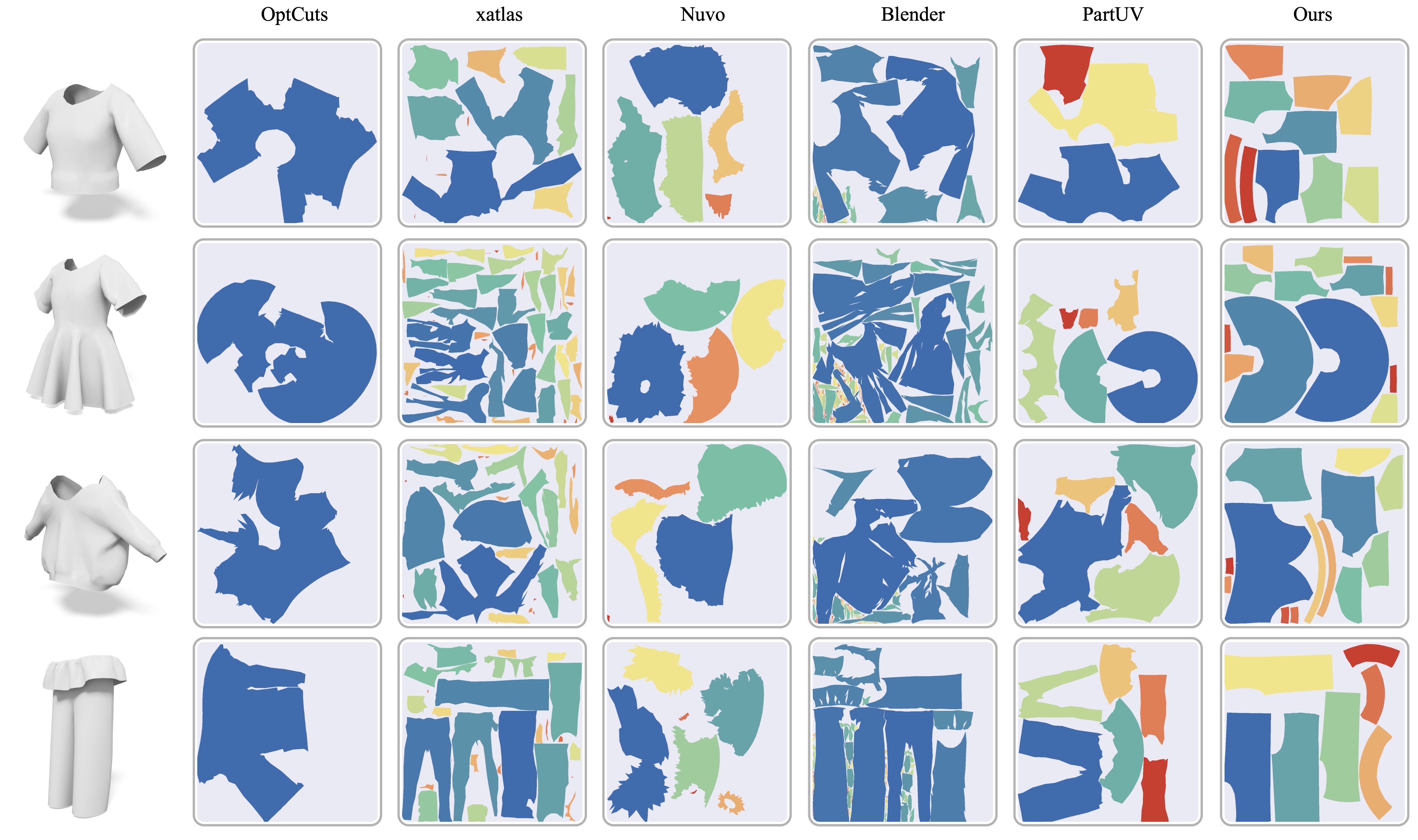}
        \caption{
            \textbf{UV layout comparison on GarmentCodeData.}
            Each island is shown with a unique color to reveal chart fragmentation and boundary regularity.
        }
        \label{fig:garment_chart}
    \end{minipage}
\end{figure*}

\begin{figure*}[p]
    \centering
    \begin{minipage}[c][\textheight][c]{\textwidth}
        \centering
        \includegraphics[width=0.97\linewidth]{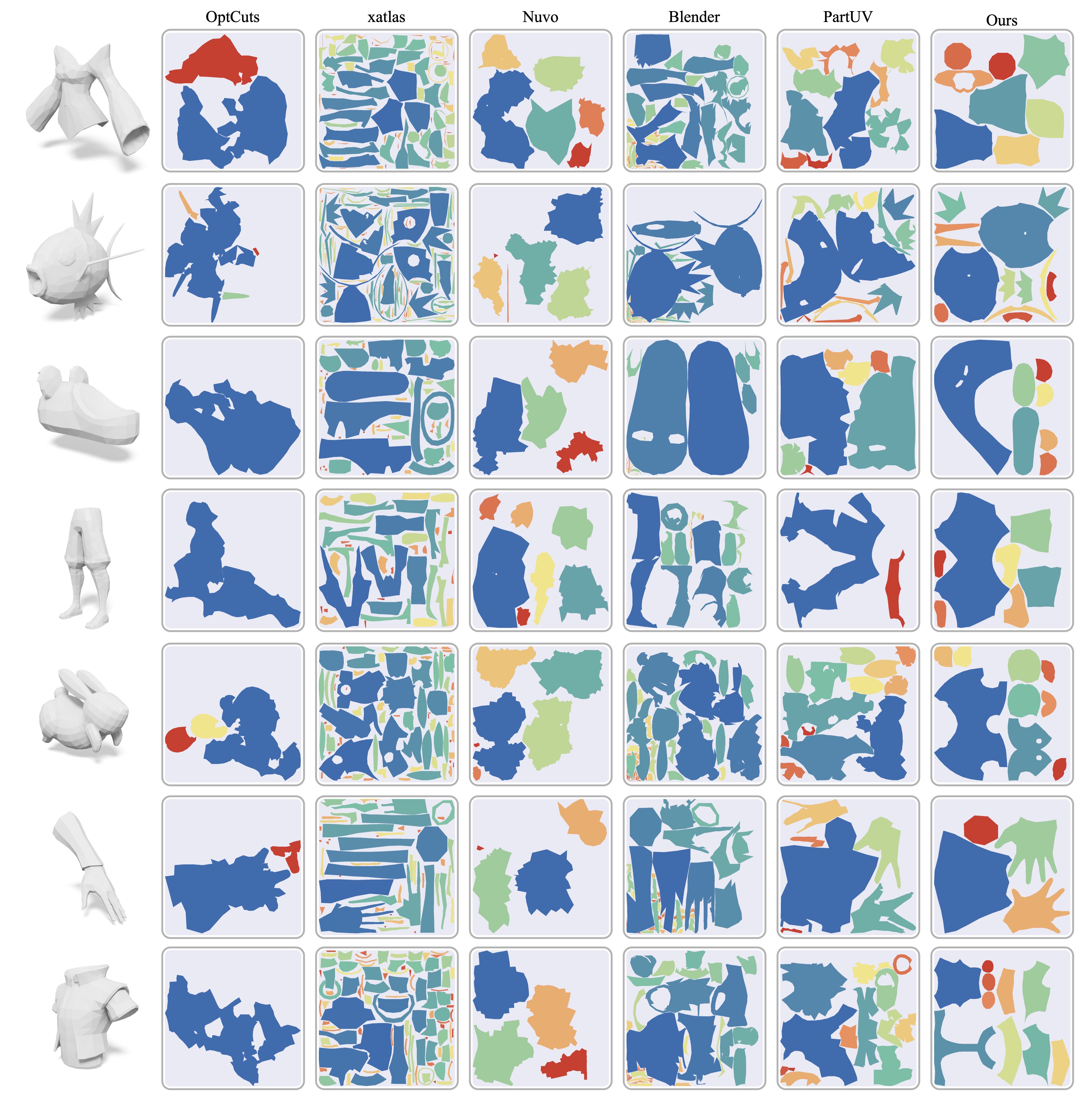}
        \caption{
            \textbf{UV layout comparison on TexVerse.}
            Each island is shown with a unique color to reveal chart fragmentation and boundary regularity.
        }
        \label{fig:texverse_chart}
    \end{minipage}
\end{figure*}

We further visualize the final \emph{UV layouts} produced by each method (Fig.~\ref{fig:garment_chart}, Fig.~\ref{fig:texverse_chart}).
Each UV island (chart) is assigned a distinct color to highlight chart count, fragmentation, and boundary regularity, while keeping the same display settings across methods.
Compared with optimization-based baselines that often yield overly fragmented atlases (many small charts) or irregular/tortuous boundaries, our method produces fewer and more coherent islands with smoother outlines.
This qualitative trend aligns with our quantitative metrics on chart complexity and boundary regularity, supporting our claim that MeshTailor layouts are easier to pack and edit in production.

\subsection{Further Analysis of Coordinate-Based Decoding}
\label{sec:supp_coord}

In the main paper, we include coordinate-based baselines (Coord-Edge/Coord-Chain) to contrast mesh-native pointer decoding with predicting seam geometry via 3D coordinates.
Here we provide a closer look at a key failure mechanism of coordinate-based decoding: the required projection (snapping) from predicted coordinates onto the input mesh.

\paragraph{Projection-induced jaggedness.}
Even when predicted segments form a visually plausible wireframe in Euclidean space, converting them into valid seams requires snapping points to mesh vertices/edges.
Because the target is a discrete mesh, small deviations in predicted coordinates can be amplified by nearest-edge snapping, producing zigzagging and jagged surface boundaries.

\paragraph{Ambiguity under nearby or interpenetrating surfaces.}
The projection step becomes particularly brittle when different surface sheets lie close to each other or interpenetrate (e.g., overlapping trouser legs).
In such cases, nearest-surface projection can become ambiguous and may collapse multiple segments onto the same sheet, yielding misplaced or disconnected seams after snapping (Fig.~\ref{fig:coord_proj}).
These artifacts are intrinsic to the projection step and motivate decoding seams directly on mesh connectivity.

\subsection{Additional Qualitative Results}
\label{sec:additional_results}
\begin{figure*}
    \centering
    \includegraphics[width=0.98\linewidth]{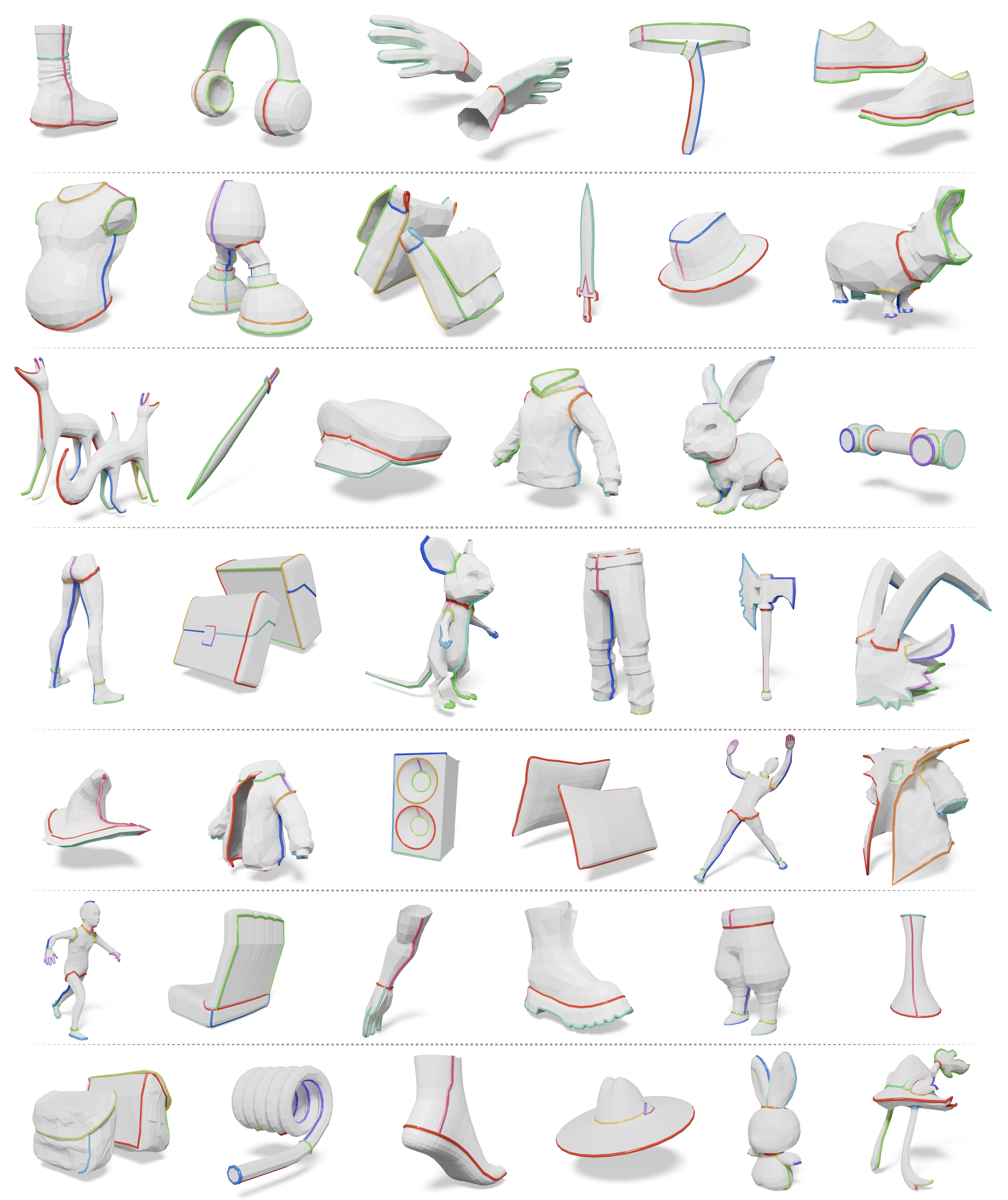}
    \caption{
    \textbf{Additional qualitative results.}
    MeshTailor produces coherent seam layouts across diverse categories and shapes.
    Colored curves denote different predicted seam chains overlaid on the input meshes.
    }
    \label{fig:gallery}
\end{figure*}

To further demonstrate the robustness and generality of MeshTailor beyond the representative examples in the main paper, we provide a gallery of additional results on diverse assets in Fig.~\ref{fig:gallery}.
The examples span a wide range of shapes and part structures, including thin components, articulated silhouettes, and complex concavities.
Across these cases, our method consistently produces clean, coherent seam layouts that follow plausible structural boundaries and avoid excessive fragmentation.

\begin{figure}[t]
  \centering
  \includegraphics[width=\linewidth]{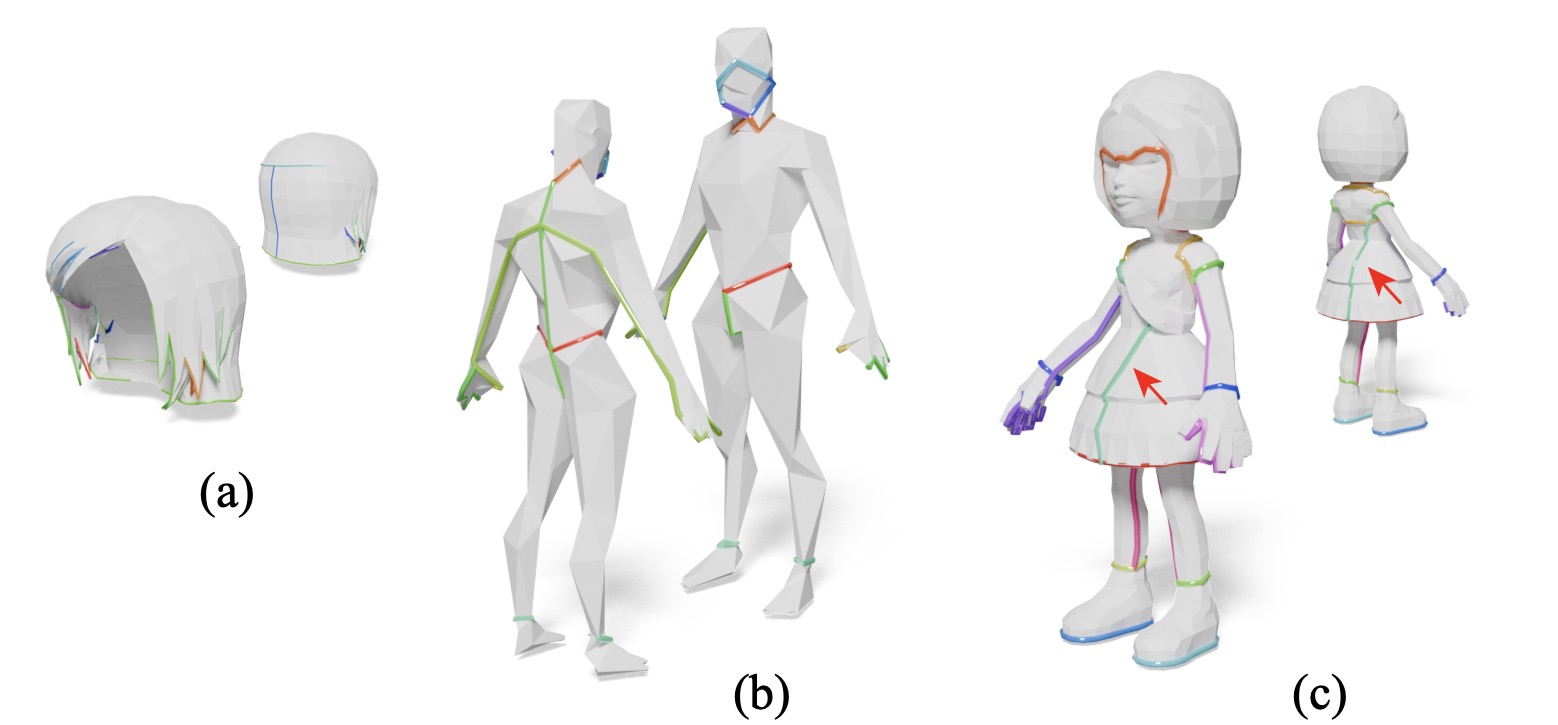}
  \caption{
    \textbf{Failure cases.}
    (a) Hair and spiky assets.
    (b) Extremely low-poly meshes.
    (c) Decoding errors can derail a chain (red arrows).
  }
  \label{fig:failure_cases}
\end{figure}

\subsection{Failure Cases and Limitations}
\label{sec:supp_failure}

Fig.~\ref{fig:failure_cases} illustrates representative failure cases and current limitations of MeshTailor.

\paragraph{Hair and highly spiky geometry.}
We find it challenging to handle hair models or assets with dense, high-frequency spikes (Fig.~\ref{fig:failure_cases}a).
Such geometry often contains thin sheets, sharp tips, and closely spaced surface layers, which makes seam placement more brittle.
Moreover, hair assets are largely out of distribution for our training data, and their seam conventions are less aligned with garment-like part boundaries.

\paragraph{Extremely low-poly meshes.}
MeshTailor may degrade on extremely low-poly meshes (Fig.~\ref{fig:failure_cases}b).
When tessellation is overly sparse, the mesh graph offers limited degrees of freedom for placing smooth, meaningful seam paths, making seams overly quantized.

\paragraph{Error propagation in stochastic decoding.}
Our decoder is probabilistic and generates seam chains autoregressively.
An incorrect vertex selection can divert the subsequent walk and corrupt the remainder of the chain (Fig.~\ref{fig:failure_cases}c).
In practice, since seams are generated as a set of chains, users can discard the problematic chain and keep the remaining valid ones.

\end{document}